\documentclass[aps,prd,preprintnumbers,nofootinbib,a4paper,11pt]{revtex4-2}
\usepackage[margin=1.1111023622in]{geometry}

\usepackage{graphicx}

\usepackage{xcolor}

\usepackage{comment}
\usepackage{url}
\usepackage{bm}
\usepackage{bbm}
\usepackage{caption}
\usepackage{feynmp-auto}
\usepackage{pifont}
\usepackage{ytableau}
\usepackage{array}
\usepackage{amsfonts}
\usepackage{amsmath}
\usepackage{amssymb}
\usepackage{physics}
\usepackage{colortbl}
\usepackage{diagbox}
\usepackage{float}

\numberwithin{equation}{section}

\usepackage{amsthm}

\usepackage[bookmarks, pagebackref=false]{hyperref}
\usepackage{color}
	\definecolor{rossoCP3}{cmyk}{0,.88,.77,.40}
		\definecolor{graa}{rgb}{0.8,0.8,0.8}
		\definecolor{blaa}{rgb}{0.2,0.2,0.6}
		\hypersetup{
			colorlinks,
			bookmarksopen,
			bookmarksnumbered,
			citecolor=blaa, 
			linkcolor=rossoCP3,	
			urlcolor=rossoCP3,	
			}
\usepackage{cleveref}

\DeclareMathOperator\Arg{Arg}
\DeclareMathOperator\sgn{sgn}

\newcommand{\beq}{\begin{eqnarray}}
\newcommand{\eeq}{\end{eqnarray}}

\newcommand{\bmp}{\noindent\begin{minipage}{16cm}}
\newcommand{\emp}{\end{minipage}\vskip 7mm} 

\def\lsim{\mathrel{\rlap{\lower4pt\hbox{\hskip1pt$\sim$}}
    \raise1pt\hbox{$<$}}}                
\def\gsim{\mathrel{\rlap{\lower4pt\hbox{\hskip1pt$\sim$}}
    \raise1pt\hbox{$>$}}}                

\begin{document}
\baselineskip=15pt
\setlength{\abovedisplayskip}{10pt}
\setlength{\belowdisplayskip}{10pt}
\setlength{\abovedisplayshortskip}{4pt}
\setlength{\belowdisplayshortskip}{4pt}
\title{\Large  Renormalization and Non-perturbative Dynamics in Conformal Quantum Mechanics} 
\author{Jacob Hafjall}\email{jahaf21@student.sdu.dk}
\author{Thomas A. Ryttov}\email{ryttov@cp3.sdu.dk}
\affiliation{{  \rm CP}$^{\bf 3}${\rm-Origins}, University of Southern Denmark, Campusvej 55, 5230 Odense M, Denmark}
\pagenumbering{arabic}

\begin{abstract}

We study conformal quantum mechanics by first considering the perturbative $S$-matrix in various dimensions. The model has two couplings and we study perturbatively the degree of ultraviolet divergences arising in the interplay between the two couplings. We then focus on the inverse square potential in one spatial dimension and compute the beta function to arbitrarily perturbative and non-perturbative orders. This we do in both the bound state sector and scattering sector. We provide explicit, exact and infinite series results of the first few non-perturbative orders.

\end{abstract}

\maketitle

\newpage

\tableofcontents

\newpage

\section{Introduction}
The regularization and renormalization of classically scale invariant mechanical systems is relevant across a wide range of areas, including condensed matter physics, atomic and molecular physics, statistical physics, and high-energy physics. The set of classically scale invariant mechanical systems has been identified by Camblong in \cite{Camblong:2000qn}. Generically, in $d$ dimensions there are two types of interactions. First, potentials of the form
\begin{align}
V(\vb{x}) \sim \frac{v(\Omega_d)}{r^2} \ ,
\end{align}
where $v(\Omega_d)$ is a dimensionless function of the $d$-dimensional solid angle are scale invariant. Second, potentials of the form
\begin{align}
V(\vb{x}) & \sim r^{d-2} \delta^{(d)}(\vb{x}) \ , \qquad d\geq 1 \ , \\
V(x) &\sim \delta'(x), \qquad d=1 \ ,
\end{align}
are also scale invariant. The first class of potentials consists of the well known \textit{inverse square potentials} (ISP)s. The initial application of the ISP was found in nuclear physics by Fermi and Teller \cite{FermiTeller1947} in a discussion on the capture of charged particles in matter and in molecular physics on discussions of low-energy electron scattering by polar molecules \cite{JeanLevy1967,fox1966wkb,fox1966variational, turner1968ground,TurnerJE1966, altshuler1957theory}. Early discussions on its Schrödinger solutions are found in \cite{MottMasseyAtomicCollisions, CASEKM1950Sp}. For a sufficiently strongly coupled ISP it was thought to be ill defined because a particle, if captured, would speed up indefinitely towards the origin, a phenomenon referred to as \textit{fall to the center} by Landau and Lifshitz \cite{landau2013quantum}.

A comprehensive review of the literature on singular potentials in quantum mechanics up until 1969 can be found in \cite{frank1971singular}. The scale symmetry of the ISP was discovered by Jackiw \cite{JACKIWR1972Iss} and later its conformal invariance in \cite{DeAlfaroV.2006Ciif}. Regularization was introduced as an attempt to make the ISP well defined past criticality. This is done by introducing a dimensionfull regulator momentarily breaking its scale invariance. However upon removing the regulator all peculiarities associated with "fall to the center" were reintroduced. It was later realized that a systematic way of consistently defining the ISP past criticality was through modern renormalization techniques which introduces running couplings to keep physics independent of the regulator. Works on the general theory of regularization and renormalization of the ISP is found in \cite{Gupta_1993, Camblong:2000ax, Camblong:2001zt, Griffiths:1DISP, Holstein,Sundaram_2021, kaplan2009conformality}. More recent work and use of the ISP is found in the renormalization of the SUSY Pöschl-Teller potential \cite{daSilva:2025vkl}, approaches to quantum gravity such as polymer quantization and functional renormalization group approaches are tested on the ISP \cite{Kunstatter:2008qx,moroz2010nonrelativistic}, in quantum cosmology on cosmological defects in a cosmic string background \cite{Bouaziz:2008wxq}, the Aharonov-Bohm effect \cite{Veloso:2024ovg,Veloso:2025slu}, on the AdS-CFT correspondance \cite{chamon2011conformal, moroz2010below}, in black hole physics \cite{srinivasan1999particle,birmingham2001near, camblong2003anomaly, burgess2018effective}, duality with the inverted harmonic oscillator \cite{Sundaram_2024}, as the interaction between a charged wire and particle \cite{denschlag1998probing, plestid2018fall}, in soft matter physics on winding transitions of floppy polymers \cite{nisoli2014attractive}, in nuclear physics through the Efimov effect \cite{efimov1973energy, fonseca1979efimov, braaten2006universality, moroz2015generalized}, in statistical physics through the Calogero-Sutherland problem \cite{calogero1969solution, sutherland1971exact} and in optics on light coherence \cite{Sundaram:16}.

The second class of scale invariant interactions includes contact terms such as the Dirac delta potential in two spatial dimensions, which has been widely studied \cite{Luty,Holstein,Mcgreevy}. This system is particularly instructive because, despite its apparent simplicity, it already exhibits a nontrivial renormalization structure. In particular reference \cite{Mcgreevy} obtained an exact beta function and showed that the model contains non-perturbative contributions in addition to the perturbative part. Related scale invariant contact interactions also arise in one dimension through $\delta'(x)$-type terms. Moreover, two-dimensional contact interactions have direct experimental relevance in ultracold atomic systems, where the anomalous breaking of scale invariance has been used to explain frequency shifts of breathing modes in two-dimensional Fermi gases \cite{Holten_2018}.

In this work we are particularly interested in computing beta functions of quantum mechanical systems to arbitrary perturbative and non-perturbative order. Computing beta functions is of central importance in quantum field theory, since they determine how couplings evolve across energy scales. Since the seminal work of Wilson \cite{Wilson:1973jj,Wilson:1974mb}, the renormalization group has provided the modern framework for understanding effective descriptions of physical systems, fixed points, universality, and dimensional transmutation. A prominent example is quantum chromodynamics (QCD), where the beta function governs the running of the strong coupling, appears in the trace anomaly, and leads to dimensional transmutation, thereby generating the scale $\Lambda_{\mathrm{QCD}}$ that sets the characteristic scale of hadronic physics. More generally, in the Standard Model coupled beta functions determine the running of gauge, Yukawa and self interactions and are essential for connecting physics across separated energy scales and for making precise phenomenological predictions at colliders and beyond.

These observations motivate the study of multi-coupling quantum mechanical systems as controlled laboratories for renormalization. Such systems retain many of the conceptual features familiar from quantum field theory, while being sufficiently simple that both perturbative and non-perturbative structures can often be exposed explicitly. They also provide a useful pedagogical setting in which the emergence of running couplings, dimensional transmutation, and the anomalous breaking of classical scale invariance can be followed in detail.

Here we initiate a study of classically scale invariant quantum mechanical systems containing multiple interactions. The motivations for this should be clear from the above discussions. While these ingredients have each been studied extensively in isolation, their combined renormalization structure has, to our knowledge, received no attention. Our aim is twofold. First, in Sec.~\ref{sec: Smatrix} we analyze the perturbative scattering matrix in various spatial dimensions and identify the divergences associated with the different interactions and their mixing. This provides a first indication of the nontrivial renormalization structure of the theory. Second, in the remainder of the paper we focus on the inverse square potential in $d=1$, where we derive the running coupling and beta function to high perturbative and non-perturbative order and show explicitly that the renormalization group flow contains a rich non-perturbative structure. We stress that by considering the inverse square potential in one spatial dimension we by no means restrict the applicability of our analysis. This is because in two or three spatial dimensions our analysis carries over to the corresponding radial Schrödinger equation. This greatly extends the work of \cite{Griffiths:1DISP}. In particular, the beta function exhibits transseries behaviour with instanton-like and multi-instanton-like contributions. In this way, the model provides both a new example of non-perturbative renormalization in quantum mechanics and a transparent setting in which such effects can be studied analytically.

\section{The Model}

In this work we want to study general scale invariant mechanical systems in $d$ spatial dimensions. We limit ourselves to Hamiltonians that do not depend explicitly on time and for $d\geq 2$ are also invariant under rotations. Hence we take $H = H(r, p)$ with $r = |{\bf x}|$ and $p = |{\bf p}|$. For a particle of mass $m$ moving in the presence of a potential the Hamiltonian is then taken to be 
\begin{align}
H = \frac{{\bf p}^2}{2m} + V(r) \ .
\end{align}
Consider now the requirement of scale invariance. If there are no parameters with dimension in the Hamiltonian the theory will be scale invariant. If we for the moment choose units $m=1$ and $\hbar =1$ then all quantities can be given the dimension of length L to some power. In particular this implies that time $t$ will have dimension $\dim t = \text{L}^2$ and the Hamiltonian will have dimension $\dim H = \text{L}^{-2}$. One could equally well have chosen energy E as the only base dimension so that the conventions would align more with relativistic quantum field theory. But we stick to length L as our base dimension. Since the potential must have dimension $\dim V = \text{L}^{-2}$ similar to the Hamiltonian and at the same time contain only dimensionless parameters there are only two possibilities left. These two terms are $\frac{1}{r^2}$ and $\frac{\delta(r)}{r}$ \cite{Camblong:2000qn,Camblong:2000ax} where $\delta(r)$ is the radial delta function. The dimension of the radial Dirac delta function is $\dim \delta(r) = \text{L}^{-1}$ so that $\dim \frac{\delta(r)}{r} = \text{L}^{-2}$ as wanted. Properties of the radial delta function will be discussed further below. Therefore the model we study in $d$ spatial dimensions has the Hamiltonian
\begin{align}\label{theory}
H = \frac{{\bf p}^2}{2m} + V(r) \ , \qquad V(r) = - \frac{c}{r^2} - k \frac{\delta(r)}{r} \ .
\end{align}
The two negative signs are chosen such that for positive couplings $c,k>0$ both terms in the potential are attractive.

A few remarks about the delta function is now in order. Consider the usual delta function $\delta^{(d)}({\bf x})$ in $d$ spatial dimensions. Its dimension is set by the requirement $\int f({\bf x}) \delta^{(d)}({\bf x}) \dd^d x =f({\bf 0})$ for some appropriate test function $f({\bf x})$ and so $\dim \delta^{(d)} ({\bf x}) = \text{L}^{-d}$.  The delta function has a number of different equivalent representations since it can be factorized in a $d$ dimensional spherical coordinate system as\footnote{Note we are considering the delta function centered at the origin. }
\begin{align}
\delta^{d} ({\bf x}) & = \frac{r^{-(d-1)}}{\Omega_d}  \delta (r) \ ,
\end{align}
where 
\begin{align}
\Omega_d = \int_{S^{d-1}} \dd \Omega_d= \frac{2\pi^{d/2}}{\Gamma(d/2)} \ ,
\end{align}
is the solid angle subtended by the full unit sphere $S^{d-1}$ in $d$ spatial dimensions. Specifically for $d=1,2,3$ it is $\Omega_1 = 2$, $\Omega_2 = 2\pi$ and $\Omega_3 = 4\pi$. The radial delta function is defined via
\begin{align}
\int_{0}^{\infty} f(r) \delta(r) \dd r &= \lim_{\epsilon \to 0 } \int_{0}^{\infty} f(r) \delta(r-\epsilon) \dd r=   \lim_{\epsilon \to 0} f(\epsilon) = f(0) \ ,
\end{align}
for some appropriate test function $f(r)$. This sets the dimension $\dim \delta(r) = \text{L}^{-1}$. A point in $d$ dimensions is specified in spherical coordinates by a radial coordinate $r$ and $d-1$ angles $\phi_i$, $i=1,\ldots,d-1$. In particular the relation between spherical coordinates and Cartesian coordinates is such that we must have ${\bf x}(r=0,\phi_i) = 0$. This obvious observation will be used in the following to check that the above representation of the delta function is indeed correct. To see this take again a suitable test function $f({\bf x})$ and the integration measure in spherical coordinates to be $r^{d-1} \dd r \dd \Omega_d$ and evaluate
\begin{align}
\int_{\mathbb{R}}  f({\bf x})  \delta^{d} ({\bf x})  \dd^dx = \int_{S^{d-1}} \int_{0}^{\infty}  f({\bf x}(r,\phi_i)) \frac{1}{\Omega_d}  \delta (r)   \dd r \dd \Omega_d  = f({\bf 0}) \frac{1}{\Omega_d} \int_{S^{d-1}}  \dd \Omega_d  =  f({\bf 0}) \ ,
\end{align}
where we first performed the radial integration before finally doing the angular integration. This shows that the delta function can be factorized in spherical coordinates as given above.  

With the factorization of the delta function in spherical coordinates in hand we can instead write the singular potential term as $ \frac{\delta(r)}{r} = 2\pi \delta^{(2)}({\bf x})$ in $d=2$ spatial dimensions. In $d=1$ what we mean is the following parity invariant term
\begin{align}
\frac{\delta(|x|)}{|x|} = \sgn x \frac{\delta(x)}{x}  =  - \sgn x \ \delta'(x) \ .
\end{align}
The Hamiltonian arrived at above is invariant under time translations and scale transformations as well as rotations for $d\geq 2$. It is also invariant under discrete parity for odd $d$ as well as time reversal. However it is of course possible to loosen some of these symmetry requirements to write other scale invariant Hamiltonians. We now briefly give two examples. For instance in $d=1$ spatial dimensions we may also at times instead consider the scale invariant but parity violating potential term
\begin{align}
V(x) & = -k' \frac{\delta(x)}{x}  = k' \delta' (x) \ .
\end{align}
For positive $k'$ this is a potential barrier on the left negative $x$-axis and a potential well on the right positive $x$-axis. It is also of great importance in $d=3$ spatial dimensions to study the dynamics of a charged particle in the presence of an electric dipole \cite{Camblong:2001zt}. In the limit of a point dipole the potential term is
\begin{align}
V(r,\theta) = \frac{qD}{4\pi \epsilon_0} \frac{\cos \theta}{r^2} \ ,
\end{align}
where $q$ is the charge of the particle and $D = |{\bf D}|$ is the magnitude of the electric dipole moment ${\bf D}$. The angle $\theta$ is the polar angle from the dipole moment ${\bf D}$ to the position ${\bf}$ of the particle ${\bf D} \cdot {\bf x} = Dr\cos \theta$. This potential is scale invariant but preserves only rotations around the axis given by the dipole moment ${\bf D}$. For a short discussion on conformal symmetry of mechanical systems in $d$ spatial dimensions and the associated conformal group see App. \ref{App: conformal symmetry}.

\section{The Perturbative $S$-matrix}\label{sec: Smatrix}
Before we dive in and investigate this conformal model in detail we first illustrate some of its peculiarities in a simple way. This will remind us of how one encounters divergences in relativistic quantum field theory. First we introduce the scattering operator
\begin{align}
S = \lim_{\substack{t_f \to \infty \\ t_i\to -\infty}}  U_{I} (t_f, t_i) \ ,
\end{align}
where 
\begin{align}
U_I(t_f,t_i)  = \text{T} e^{- \frac{i}{\hbar} \int_{t_i}^{t_f} V_I (t) dt} \ ,
\end{align}
is the time evolution operator in the interaction picture and 
\begin{align}
V_I(t) = e^{\frac{i}{\hbar} H_0 t } V e^{- \frac{i}{\hbar} H_0 t} \ , \qquad H_0  = \frac{{\bf p}^2}{2m} \ ,
\end{align}
is the potential in the interaction picture. Consider now the matrix elements of the scattering operator (the $S$-matrix) in momentum eigenstates. It has the following expression
\begin{align}
\langle {\bf p}_f | S | {\bf p}_i \rangle & = \delta^d({\bf p}_f - {\bf p}_i )  + 2 \pi i  \delta (E_f - E_i ) \langle {\bf p}_f | T | {\bf p}_i \rangle  \ ,
\end{align}
with the matrix elements of the transition operator (the $T$-matrix) having the following expansion
\begin{align}
 \langle {\bf p}_f | T | {\bf p}_i \rangle  & =  \langle {\bf p}_f | V | {\bf p}_i \rangle + \int d^d p \langle {\bf p}_f | V | {\bf p} \rangle \frac{1}{E_i - \frac{{\bf p}^2}{2m} + i \epsilon } \langle {\bf p} | V |  {\bf p}_i \rangle + \ldots  \ .
\end{align}
We now observe the divergences that begin to appear in the transition matrix order by order. First we have to calculate the matrix element $\langle {\bf p'} | V | {\bf p} \rangle$ since it enters in all of the terms at each order. This we do in App. \ref{Fourier}.

\subsection{$d=1$}
We begin in $d=1$ spatial dimensions where $r= |x|$. At first order the transition matrix is
\begin{align}
\langle p_f | \left( \frac{-c}{x^2} - k \frac{\delta(|x|)}{|x|} \right) | p_i \rangle & = \frac{c}{2\hbar^2}  |p_f-p_i| - \frac{k}{4\pi \hbar^2} \Lambda \ .
\end{align}
At first order there is no divergence in the ISP term but there is a linear divergence $\sim \Lambda$ in the Dirac singular term. At second order there are three terms that are proportional to $c^2$, $k^2$ and $ck$ respectively. Consider first the $c^2$ term
\begin{align}
\left( \frac{c}{2 \hbar^2} \right)^2 \int_{-\infty}^{\infty} \frac{|p_f - p| |p - p_i| }{E_i - \frac{p^2}{2m} + i \epsilon} \dd p= \left( \frac{c}{2 \hbar^2} \right)^2 \int_{-\infty}^{\infty}  \frac{|p_f||p_i|- \frac{|p_f| |p_i| (p_f+p_i)}{p_fp_i} p + \frac{|p_f||p_i|}{p_fp_i} p^2}{E_i - \frac{p^2}{2m} + i \epsilon} \dd p \ .
\end{align}
The numerator in the integrand has three terms each proportional to $p^0$, $p^1$ and $p^2$. The integral over the term proportional to $p^0$ is finite. The integral over the term proportional to $p^1$ vanishes since the integrand is odd. The term that goes as $p^2$ in the numerator however is problematic and diverges at large momenta 
\begin{align}
\sim \left( \frac{c}{2 \hbar^2} \right)^2 \int^{\Lambda}  \frac{p^2}{p^2 + i \epsilon } dp \sim \Lambda  \ .
\end{align}
So the transition matrix at second order for the $c^2$ term has a linear divergence $\sim \Lambda$. Consider now the $k^2$ term at second order in the transition matrix
\begin{align}
\left( \frac{k}{4\pi\hbar^2} \right)^2 \Lambda^2 \int_{-\infty}^{\infty} \frac{1}{E_i - \frac{{\bf p}^2}{2m} + i \epsilon} \dd p \sim \Lambda^2 \ .
\end{align} 
We see it has a quadratic divergence since the integral is finite. At last consider the mixed term $ck$ between the ISP and the Dirac singular term for which the transition matrix at second order is
\begin{align}
\frac{c}{2 \hbar^2} \frac{-k}{4\pi\hbar^2} \Lambda \int_{-\infty}^{\infty} \frac{|p_f-p| + |p-p_i| }{E_i - \frac{{\bf p}^2}{2m} + i \epsilon}  \dd p = \frac{c}{2 \hbar^2} \frac{-k}{4\pi\hbar^2} \Lambda \int_{-\infty}^{\infty} \frac{ |p_f| + |p_i| - \left( \frac{|p_f|}{p_f} + \frac{|p_i|}{p_i} \right) p   }{E_i - \frac{{\bf p}^2}{2m} + i \epsilon}  \dd p \sim \Lambda \ .
\end{align}
The numerator has a term proportional to $p^0$ and $p^1$. The integral over the first term is finite while the second vanishes since the integrand is odd. Hence the divergence of the mixed term is linear $\sim \Lambda$ at second order. 

To complete the $d=1$ story we also consider the parity odd potential term with coupling $k'$ for which the first term in the transition matrix is
\begin{align}
\langle p_f | \frac{-k' \delta (x)}{x} | p_i \rangle  =  i \frac{k'}{2\pi \hbar} \frac{p_f-p_i}{\hbar} \ .
\end{align}
We see it is finite similar to the $c$ term above. Also similar to the $c^2$ term above is the appearance of a linear divergence $\sim \Lambda$  at second order
\begin{align}
\left( i \frac{k'}{2\pi \hbar} \right)^2 \int_{-\infty}^{\infty} \frac{(p_f-p)(p-p_i)}{E_i - \frac{{\bf p}^2}{2m} + i \epsilon} \dd p \sim \Lambda \ .
\end{align}
The mixed term $ck'$ is 
\begin{align}
\frac{c}{2\hbar^2} \frac{i k'}{2\pi \hbar} \int_{-\infty}^{\infty} \frac{|p_f-p|(p-p_i) + (p_f-p)|p-p_i|}{E_i - \frac{p^2}{2m} + i \epsilon} \sim \Lambda \ .
\end{align}
We conclude that in $d=1$ spatial dimensions all three couplings $c,k$ and $k'$ exhibit a divergence in the transition matrix. For the $c$ and $k'$ couplings it appears at second order in the transition matrix while for the $k$ coupling it appears at first order. In the mixed term $ck$ there is also a linear divergence at second order. We summarize the results in the table below
\begin{table}[h]
\begin{tabular}{|l|l|l|l|l|l|l|l|l|}
\hline
Term       & $c$ & $k$       & $k'$ & $c^2$     & $k^2$       & $k'^2$    & $ck$      & $ck'$     \\ \hline
Divergence & $1$ & $\Lambda$ & $1$  & $\Lambda$ & $\Lambda^2$ & $\Lambda$ & $\Lambda$ & $\Lambda$ \\ \hline
\end{tabular}
\caption{Degree of divergence in $d=1$ in the transition matrix at first and second order.}
\end{table}

\subsection{$d=2$}
At first order the transition matrix is
\begin{align}
\langle {\bf p}_f |\left(  \frac{-c}{r^2} - k \frac{ \delta(r)}{r} \right) | {\bf p}_i \rangle =  - \frac{c}{2\pi\hbar^2} \ln  \frac{\Lambda}{|{\bf p}_f-{\bf p}_i|}  - \frac{k}{2\pi \hbar^2} + O(\Lambda^{-2}) + \text{finite terms} \ .
\end{align}
We see that the $c$ term has a logarithmic divergence while the $k$ term has no divergence at first order. From now we do not write the terms that are zero or finite as we remove the cutoff. Consider the second order in the transition matrix. The $c^2$ term is
\begin{align}
\left( - \frac{c}{2\pi\hbar^2} \right)^2 \int_{\mathbb{R}^2} \frac{ \ln  \frac{\Lambda}{|{\bf p}_f-{\bf p}|} \ln  \frac{\Lambda}{|{\bf p}-{\bf p}_i|} }{E_i - \frac{{\bf p}^2}{2m} +  i \epsilon} \dd^2 p \ .
\end{align}
At large $|{\bf p}|$ the logarithm terms go as $\sim \ln \frac{\Lambda}{|{\bf p}|}$ and so the transition matrix goes as
\begin{align}
\left( - \frac{c}{2\pi\hbar^2} \right)^2 \int^{\Lambda} \frac{\left( \ln \frac{\Lambda}{|{\bf p}|} \right)^2}{|{\bf p}|} \dd |{\bf p}| \sim \Lambda^0  \ .
\end{align}
We see there is no divergence in the $c^2$ term. Consider now the $k^2$ term which is logarithmically divergent 
\begin{align}
\left(  - \frac{k}{2\pi \hbar^2} \right)^2 \int \frac{1}{E_i - \frac{{\bf p}^2}{2m} + i \epsilon}  \dd^2 p \sim  \int^{\Lambda} \frac{ 1 }{|{\bf p}|} \dd |{\bf p}| \sim \int^{\frac{\Lambda}{|{\bf p}_f-{\bf p}_i|}} \frac{1}{\mu} \dd \mu \sim \ln \frac{\Lambda}{|{\bf p}_f - {\bf p}_i|} \ .
\end{align}
Thus the transition matrix at second order also has a logarithmic divergence in the $k^2$ term similar to the $c$ term at first order. At last consider the mixed term $ck$
\begin{align}
\left( - \frac{c}{2\pi\hbar^2} \right) \left(- \frac{k}{2\pi \hbar^2} \right) \int_{\mathbb{R}^2} \frac{ \ln  \frac{\Lambda}{|{\bf p}_f-{\bf p}|} +  \ln  \frac{\Lambda}{|{\bf p}-{\bf p}_i|} }{E_i - \frac{{\bf p}^2}{2m} +  i \epsilon} \dd^2 p \ .
\end{align}
At large $|{\bf p}|$ the logarithm terms go as $\sim \ln \frac{\Lambda}{|{\bf p}|}$ and so the transition matrix goes as
\begin{align}
\left( - \frac{c}{2\pi\hbar^2} \right) \left(- \frac{k}{2\pi \hbar^2} \right)\int^{\Lambda} \frac{  \ln \frac{\Lambda}{|{\bf p}|}}{|{\bf p}|} \dd |{\bf p}| \sim \Lambda^0 \ .
\end{align}
So there is no divergence in the mixed $ck$ term. 
\begin{table}[h]
\begin{tabular}{|l|l|l|l|l|l|}
\hline
Term       & $c$           & $k$ & $ck$ & $c^2$ & $k^2$         \\ \hline
Divergence & $\ln \Lambda$ & $1$ & $1$  & $1$   & $\ln \Lambda$ \\ \hline
\end{tabular}
\caption{Degree of divergence in $d=2$ in the transition matrix at first and second order.}
\end{table}

\subsection{$d \geq 3$}
We now consider the situation in $d\geq 3$ spatial dimensions where we only have to consider the contributions from the coupling $c$ since the Dirac delta singular potential vanishes. In $d\geq3$ we find for the transition matrix at first order
\begin{align}
 \langle {\bf p}_f | \frac{-c}{r^2} | {\bf p}_i \rangle & =   \frac{ - c}{(d-2) \hbar^2 \Omega_d} \frac{1}{|{\bf p}_f-{\bf p}_i|^{d-2}} \ .
 \end{align}
At second order the transition matrix is
\begin{align}
\left( \frac{-c}{(d-2) \hbar^2 \Omega^2} \right)^2 \int \frac{1}{|{\bf p}_f - {\bf p} |^{d-2} \left( E_i  - \frac{{\bf p}^2}{2m} + i \epsilon \right)|{\bf p}-{\bf p}_i|^{d-2}} \dd^d p \ .
\end{align}
At large $|{\bf p}|$ the numerator goes as $|{\bf p}|^{d-1}$ from the integration measure in spherical coordinates while the denominator goes as $|{\bf p}|^{d-2} |{\bf p}|^{2} |{\bf p}|^{d-2} = |{\bf p}|^{2(d-1)}$. Therefore the integrand goes as $| {\bf p}|^{1-d}$ at large $|{\bf p}|$ and there is no ultraviolet divergence for any $d \geq 3$ to this order.

\section{Inverse Square Potential on the Positive Axis}\label{posISP}
We consider the spectrum of a particle of mass $m$ in a $1d$ ISP constrained to the positive axis with
\begin{equation}
V(x) = \left\{ \begin{array}{rl}
\infty, & x \leq 0 \ ,\\
-\frac{c}{x^2}, & x > 0 \ ,
\end{array} \right.
\end{equation}
where $c>0$ so that the potential is attractive. It is well known, that when $2m c/\hbar^2 < 1/4$ the potential is too weak for bound states to exist. This is seen by the absence of normalizable solutions. When $2m c/\hbar^2 > 1/4$, a standard textbook treatment lead to inconsistent behaviour reflecting the fact that an appropriate regularization and renormalization scheme is needed to extract physical relevant information of the model. As we are interested in the critical regime we will restrict ourselves to when $2m c/\hbar^2 > 1/4$ and for this purpose rewrite the coupling as $g^2 = 2m c/\hbar^2 - 1/4$, with $g \geq 0$. Then, we regularize the potential such that the wall at $x=0$ is moved to $x=\epsilon$ and consequently, wave functions are subject to the Dirichlet boundary condition
\begin{align}\label{eq: DBC}
	\psi(\epsilon) = 0 \ .
\end{align}

\subsection{General Solution}
For $x > \epsilon$ the time-independent Schrödinger wave equation is
\begin{align}
\label{eq: Schrodinger wave equations}
\left( \frac{d^2}{dx^2} + \frac{g^2 + \frac{1}{4}}{x^2} + k^2  \right) \psi(x) &= 0 \ , &k^2& = \frac{2m}{\hbar^2} E \ , &E&>0 \\
\label{eq: schrödinger bound equation}
\left( \frac{d^2}{dx^2} + \frac{g^2 + \frac{1}{4}}{x^2}    - \kappa^2  \right) \psi(x) &= 0 \ , &\kappa^2& =  -  \frac{2m}{\hbar^2} E \ , &E&<0
\end{align}
As a solution ansatz, we consider
\begin{align}
\label{scatteringansatz}
\psi_k(x) &= (kx)^{\frac{1}{2}} F(kx)  \ , &E&>0  \ ,\\
\label{boundansatz}
\psi_{\kappa}(x) &= (\kappa x)^{\frac{1}{2}} G (\kappa x)  \ , &E&<0 \ ,
\end{align}
leading to a differential equation for $F$ and $G$ of the form
\begin{align} 
(k x)^2 \frac{d^2 F (k x) }{d (k x)^2}  + (k x) \frac{d F (k x)}{d (k x)}  + \left[ (k x)^2 - (ig)^2\right] F( k x)  & = 0 \ , &E&>0 \ . \\
(\kappa x)^2 \frac{d^2 G (\kappa x) }{d (\kappa x)^2}  + (\kappa x) \frac{d G (\kappa x)}{d (\kappa x)}  - \left[ (\kappa x)^2 + (ig)^2\right] G( \kappa x) & = 0 \ , &E& < 0 \ .
\end{align}
Thus $F$ satisfies Bessel's differential equation and $G $ satisfies the modified Bessel's differential equation (See Eq.s \eqref{eq: bessels equation} and \eqref{eq: modified bessels equation} in App. \ref{App: bessel}). Since these are second order linear differential equations there are two linearly independent solutions in each case. We denote the two linearly independent solutions to Bessel's equations as $J_{ig}$ and $Y_{ig}$ while the two linearly independent solutions to the modified Bessel's equations are denoted as $I_{ig}$ and $K_{ig}$. The general solutions to the time-independent Schrödinger wave equation are therefore a linear combination of Bessel and modified Bessel functions in the following way
\begin{align}
\label{eq:sol1}
\psi_k(x) & = (kx)^{\frac{1}{2}} \left[ A J_{ig} (kx) + B Y_{ig}(kx) \right]  \ , &E&>0 \ . \\
\label{eq:sol2}
\psi_{\kappa}(x) & = (\kappa x)^{\frac{1}{2}}\left[ C I_{ig} (\kappa x) + D K_{ig}(\kappa x) \right] \ , &E&<0 \ ,
\end{align}
where $A,B,C,D$ are constants.

\subsection{Bound State Sector $E<0$}\label{sec: bound sector}
The modified Bessel function of the first kind $I_{ig}(\kappa x)$ is exponentially growing on the positive $x$-axis and therefore has to be discarded since it cannot be normalized. The modified Bessel function of the second kind $K_{ig}(\kappa x)$, on the other hand, is exponentially decaying on the positive $x$-axis and hence can be normalized \cite{Table_of_Integrals_Series_Products}. The bound state solution on the positive $x$-axis therefore is  
\begin{align}\label{eq:boundstate}
\psi_{\kappa}(x) & = D_{\epsilon} (\kappa x)^{\frac{1}{2}} K_{ig}(\kappa x) \ , &E&<0 .
\end{align}
The normalization of the bound state solution of course depends on $\epsilon$ and we have therefore chosen to write it as a subscript on $D_{\epsilon}$. We have chosen not to write the explicit result for the integral since the explicit value for $D_{\epsilon}$ is irrelevant. By writing $K_{ig}$ in terms of $I_{\pm ig}$ (See App. \ref{App: bessel} for useful relations on Bessel functions), we obtain
\begin{align}\label{eq: solsolsol}
\psi_{\kappa} (x) = - D_{\epsilon} \frac{\pi}{\sinh (\pi g)}  (\kappa x)^{\frac{1}{2}} \left| I_{ig}(\kappa x) \right| \sin \left[ \arg I_{ig}(\kappa x) \right] \ .
\end{align}
Imposing the boundary condition Eq. \eqref{eq: DBC} 
\begin{align}
	- D_{\epsilon} \frac{\pi}{\sinh (\pi g)}  (\kappa \epsilon)^{\frac{1}{2}} \left| I_{ig}(\kappa \epsilon) \right| \sin \left[ \arg I_{ig}(\kappa \epsilon) \right] =  0 \ ,
\end{align}
leading to
\begin{align}
	\sin[\arg I_{ig}(\kappa \epsilon)] = 0 \ .
\end{align}
This serves as the quantization condition and gives
\begin{align}
\label{eq: bound state argI condition}
 \arg I_{ig}(\kappa_n \epsilon) + n \pi = 0 \ , \qquad n=0,\pm 1, \pm 2,\ldots \ .
\end{align}
Of course, this is equivalent to imposing the boundary condition $\psi_{\kappa} (\epsilon) = 0$. To elucidate the non-perturbative features of the model, it is convenient to rewrite the multi-valued argument as follows. Starting with 
\begin{align}
	I_{ i g }(\kappa x) &= \left( \frac{\kappa x}{2} \right)^{ig}  \tilde{I}_{ig}(\kappa x)  \ , \qquad \tilde{I}_{ig}(\kappa x) =  \sum_{m=0}^{\infty} \frac{1}{ m! \Gamma (m+ig +1)} \left( \frac{\kappa x}{2} \right)^{2m} \ ,
\end{align}
we note that
\begin{align}
\arg I_{ig} (\kappa_{n} \epsilon)  = g \ln \frac{\kappa_{n} \epsilon}{2} + \Arg \tilde{I}_{ig} (\kappa_{n} \epsilon) + 2\pi k \ , \qquad k=0,\pm1,\pm2,\ldots \ ,
\end{align}
and therefore, the quantization condition is equivalently written as
\begin{align}\label{eq:QCn}
g \ln \frac{\kappa_{n} \epsilon}{2} + \Arg \tilde{I}_{ig} (\kappa_{n} \epsilon) + (2 k + n ) \pi = 0  \ ,
\end{align}
where $\Arg$ denotes the principal argument function defined in the conventional way $-\pi < \Arg z \le \pi$. Due to the multi-valued nature of the argument this quantization condition now contains an integral factor $k$. We will soon see that the factor $k$ should be understood as parametrizing a multivalued running coupling $g_k$ where each $k$ defines a particular branch of the running coupling. We will also see further down that the ground state can be chosen to be the $n=1$ state and that all states with $n>1$ are exited states. States with $n<0$ should be discarded. So for now we will just consider the ground state $n=1$ and discuss the fate of all other states later on. The quantization condition for the ground state is
\begin{align}
g \ln \frac{\kappa_1 \epsilon}{2} + \Arg \tilde{I}_{ig} (\kappa_1 \epsilon) +  (2k+1)\pi = 0 \ .
\end{align}
At this point we will switch from working with a small distance cutoff $\epsilon$ to instead working with a large momentum cutoff
 \begin{align}
 \Lambda = \frac{2 \hbar}{\epsilon} \ .
 \end{align}
Similarly, instead of working with the ground state energy $E_1 = - \frac{(\hbar \kappa_1)^2}{2m}$, or equivalently $\kappa_1$, we will work with
 \begin{align}
 \Lambda_{IR} = \hbar \kappa_{1}  \ , \qquad E_1 = -  \frac{\Lambda_{IR}^2}{2m} \ .
 \end{align}
The low scale $\Lambda_{IR}$ is set by the ground state energy and is measured in units of momentum.  The factor of two in the definition of the cutoff $\Lambda$ is purely conventional. These conventions fit more naturally with the ones in relativistic quantum field theory and particle physics. For the ground state, the quantization condition then is
\begin{align}\label{eq: QCgroundstate}
 g \ln \frac{\Lambda_{IR}}{\Lambda} + \Arg \tilde{I}_{ig} \left( \frac{2\Lambda_{IR}}{\Lambda} \right)  + (2k+1)\pi = 0 \ .
\end{align}

\subsubsection{The Ground State}
The next natural step to take would be to solve for the ground state $\Lambda_{IR}$ for a given cutoff $\Lambda$ and value of the coupling $g$.  We will write down a solution in terms of a transseries which in principle can be worked out to arbitrary order. First we note that the ground state quantization condition in Eq. \eqref{eq: QCgroundstate} depends only on the ratio $\frac{\Lambda_{IR}}{\Lambda}$ and so the solution for the ground state must be of the form
\begin{align}
\Lambda_{IR} = \Lambda f(g) \ , 
\end{align}
for some function $f(g)$ of the coupling. Unfortunately, the ground state quantization condition that we have to solve for $\frac{\Lambda_{IR}}{\Lambda}$ does not lend itself to an exact solution and so we have to resort to an approximation of some sort. We will begin by first appropriately writing the ground state quantization condition as a power series in $\frac{\Lambda_{IR}}{\Lambda}$ with real powers. But here we have to be a bit careful  since the logarithm term $\ln \frac{\Lambda_{IR}}{\Lambda}$ is of course non-analytic at zero. 
To deal with this term we exponentiate the ground state quantization condition in Eq. \eqref{eq: QCgroundstate} to get
\begin{align}
\frac{\Lambda_{IR}}{\Lambda} e^{   \frac{1}{g} \Arg \tilde{I}_{ig} \left( \frac{2\Lambda_{IR}}{\Lambda} \right)   } - e^{  - \frac{(2k+1)\pi}{g} }= 0 \ .
\end{align}
The next step is now to investigate how $\Arg \tilde{I}_{ig}$ depends on $\Lambda_{IR}/ \Lambda$. We can use the identity $\Gamma(1+z) = z \Gamma(z)$ repeatedly to write
\begin{align}
\tilde{I}_{ig}  \left( \frac{2\Lambda_{IR}}{\Lambda} \right) &= \frac{1}{\Gamma(1+ig)} \eta_{ig} \left(\frac{ \Lambda_{IR}}{\Lambda}\right), \\
\eta_{ig}  \left( \frac{\Lambda_{IR}}{\Lambda} \right) &\equiv 1 + \sum_{m=1}^\infty \frac{1}{m!} \left(\prod_{n=0}^{m-1} \frac{1}{1+ig+n}\right) \left(\frac{\Lambda_{IR}}{\Lambda}\right)^{2m} \ .
\end{align}
Then
\begin{align}
\Arg \tilde{I}_{ig}  = - \Arg \Gamma(1+ig) + \Arg \eta_{ig} \ .
\end{align}
The second term on the right-hand-side $\Arg \eta_{ig}$ can be written as a power series in even powers of $\Lambda_{IR}/\Lambda$ starting from second order. We can therefore write the ground state quantization condition as 
\begin{align}
 - e^{\frac{1}{g} \left(\Arg \Gamma(1+ig) - (2k+1)\pi \right)} + \frac{\Lambda_{IR}}{\Lambda} e^{\frac{1}{g} \Arg \eta_{ig}} = 0 \ .
\end{align}
Since $\Arg \eta_{ig}$ can be written as a power series in $\Lambda_{IR} / \Lambda$ with even powers so can $e^{\frac{1}{g} \Arg \eta_{ig}}$. We end up writing the ground state quantization condition as
\begin{align}
\label{eq: 1D_ISP_bound_state_poly}
a_0 e^{- \frac{(2k+1)\pi}{g}}+ \sum_{i=0}^{\infty} a_{2i+1} \left(\frac{\Lambda_{IR}}{\Lambda} \right)^{2i+1} = 0  \ ,
\end{align}
where 
\begin{align}
a_0 &=- e^{  \frac{1}{g}  \Arg  \Gamma(1+ig)  } = e^{-\gamma} \left(   -1 + \frac{1}{6} \psi^{(2)}(1) g^2 + \ldots  \right) \ , \\
a_1 &=1  \ , \\
a_3 &= - \frac{1}{1+g^2} \ , \\
a_5 &= \frac{9}{2(1+g^2)^2(4+g^2)}  \ , \\
a_7 &= \frac{-263 + 95g^2  - 2g^4}{6(1+g^2)^3 (36 +13g^2 +g^4)} \ , \\
 & \vdots \nonumber
\end{align}
A couple of comments are now in order. We want to find a solution for $\frac{\Lambda_{IR} }{ \Lambda }$ but what should our ansatz for a solution look like? Note the appearance of the term $e^{-  \frac{(2k+1) \pi}{g} }$ which is non-analytic at $g=0$ and reminds us of non-perturbative instantonic physics. This exponentially suppressed term is bound to appear in the solution for the ground state. We will therefore solve for the ground state as a series in this non-perturbative term. Also note that since all the coefficients $a_0$ and $a_{2i+1}$ are power series in $g^2$ the coefficients of the non-analytic terms will be power series in $g^2$. With these observations in mind we take the following  ansatz
\begin{align}\label{eq: f ansatz1}
\frac{\Lambda_{IR}}{\Lambda} &=-a_0 e^{-  \frac{(2k+1) \pi}{g}} + \sum_{l=1}^{\infty} f_{2l+1} e^{-(2l+1)  \frac{(2k+1) \pi}{g}}  \nonumber \\
&= - a_0  e^{-  \frac{(2k+1)\pi}{g} } + f_3  e^{-3  \frac{(2k+1) \pi}{g}} +  f_5  e^{-5 \frac{(2k+1) \pi}{g}} + f_7  e^{-7 \frac{(2k+1) \pi}{g}}  + \ldots \ ,
\end{align}
as our ansatz for a solution. The solution vanishes for vanishing coupling as it should. We now show how to fix the solution unambiguously and uniquely by fixing the coefficients $f_{2l+1}$ order by order. Note that there are no logarithmic terms in our ansatz. These are not needed to uniquely fix a solution. Plugging the ansatz in Eq. \eqref{eq: f ansatz1} into the ground state quantization condition as written in Eq. \eqref{eq: 1D_ISP_bound_state_poly}, and expanding in $e^{- \frac{(2k+1) \pi}{g}}$ we find
\begin{align}
0 &=  \left(-  a_0^3 a_3 + f_3 \right) e^{-3  \frac{(2k+1) \pi}{g}} + \left( - a_0^5 a_5+ 3 a_0^2  a_3 f_3 + f_5 \right) e^{-5  \frac{(2k+1) \pi}{g} } \nonumber \\
&~  + \left( - a_0^7 a_7 + 5 a_0^4  a_5 f_3 - 3 a_0 a_3f_3^2  +3 a_0^2 a_3 f_5 + f_7 \right) e^{-7  \frac{(2k+1) \pi}{g} } + \ldots \ .
\end{align}
The non-analytic exponential terms can be considered as independent from the analytic powers $g^m$ since one cannot expand one in terms of the other. Therefore, this can only be satisfied provided each term vanishes. This gives
\begin{align}
f_3 &=a_0^3 a_3 \ , \\
f_5 &= a_0^5 \left( - 3 a_3^2 + a_5 \right) \ , \\
f_7 &=  a_0^7 \left( 12a_3^3 - 8 a_3a_5 + a_7 \right) \ , \\
\vdots \nonumber 
\end{align}
and so at a given cutoff $\Lambda$ and coupling $g$ the ground state is
\begin{align}\label{groundstate}
\Lambda_{IR} &=\Lambda f(g) \ , 
\end{align}
with 
\begin{align}
\label{eq: bound sector f result}
f(g) & = e^{\frac{1}{g} \Arg \Gamma(1+ig) } e^{- \frac{(2k+1) \pi}{g}}  + \frac{ e^{\frac{3}{g} \Arg \Gamma(1+ig)} }{1+g^2} e^{- 3 \frac{(2k+1) \pi}{g}}  + \frac{ 3(5+2g^2) e^{\frac{5}{g} \Arg \Gamma(1+ig)} }{2(1+g^2)^2(4+g^2)} e^{- 5 \frac{(2k+1) \pi}{g}}   \nonumber \\
\quad & + \frac{(911+625g^2+74g^4) e^{\frac{7}{g} \Arg \Gamma(1+ig)} }{6(1+g^2)^3(4+g^2)(9+g^2)}  e^{ -7 \frac{(2k+1) \pi}{g}} + \ldots \ .
\end{align}
We stress that every coefficient to the non-analytic term $e^{- \frac{(2k+1) \pi}{g}}$ is an exact power series in the coupling $g$. This is an important result and we will make important use of it further into the text when we calculate the beta function of the theory. If we expand every power series coefficient it can equally well be written in the usual transseries format
\begin{align}
f(g) & = \left[ 1- \frac{1}{6} \psi^{(2)}(1) g^2 + \ldots  \right] e^{- \left( \frac{(2k+1) \pi}{g} + \gamma \right) } \nonumber \\
& + \left[1 - \frac{1}{2}\left( 2+ \psi^{(2)}(1) \right) g^2  + \ldots \right] e^{- 3 \left(\frac{(2k+1) \pi}{g} + \gamma\right)} \nonumber \\
& + \left[ \frac{15}{8} -\frac{1}{32} \left( 111 + 50 \psi^{(2)}(1)  \right) g^2 + \ldots  \right] e^{- 5 \left( \frac{(2k+1) \pi}{g} + \gamma\right) } \nonumber \\
& + \left[ \frac{911}{216}  -\frac{7}{7776} \left( 12533 + 5466 \psi^{(2)}(1) \right) g^2 + \ldots  \right] e^{- 7 \left( \frac{(2k+1) \pi}{g} + \gamma\right) } \\
& ~ \vdots \nonumber
\end{align}

\subsubsection{The Running Coupling}

Now follows a crucial observation. The existence of the ground state is a feature of the system we want to maintain and therefore to \emph{keep it fixed as we vary the cutoff $\Lambda$}. In other words, we want to keep $\Lambda_{IR}$ fixed and independent of the cutoff $\Lambda$. One can imagine that the ground state energy $- \frac{\Lambda_{IR}^2}{2m}$ is something we have measured in an experiment. Importantly, we then want to express all other physical quantities and observables in terms of this. Following, this strategy then clearly implies that the coupling $g$ \emph{must become dependent on the cutoff $g = g(\Lambda)$},  so that as we vary the cutoff the ground state energy does not change and stays fixed. This cutoff dependent coupling $g(\Lambda)$ is called the \emph{bare coupling}.  As the bare coupling changes as we vary the cutoff, we say that it \emph{runs}. The running coupling $g(\Lambda)$ at each value of $\Lambda$ defines a family of effective theories with different cutoffs but with the low-energy physics (ground state energy, scattering phase shift, etc) being the same.

In practice what we will do is to take a theory where physical observables are originally expressed in terms of a dimensionless coupling $g$ and then swap this, so that all physical observables will instead be expressed in terms of a dimensionful measured quantity $\Lambda_{IR}$. This phenomenon is an example of dimensional transmutation \cite{coleman1973radiative}.  All physical observables must be independent of the cutoff $\Lambda$. This implies for instance that the coupling $g(\Lambda)$ as defined above cannot be a physical observable.

\begin{figure}[t]
    \centering
    \includegraphics[width=.6\linewidth]{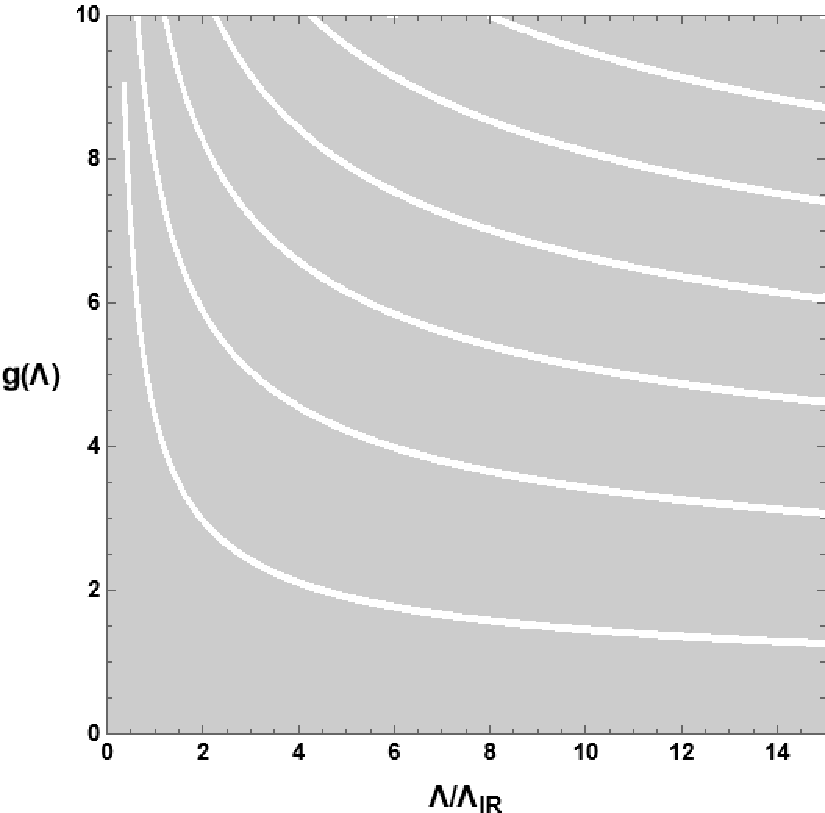}
    \caption{The white curves are the  multivalued running coupling for $k=0,1,2,3,4,5$ determined by the running coupling condition Eq. \eqref{eq:runningcoupling}. Note that the coupling goes to zero as we remove the cutoff. }
    \label{fig: 1D_ISP_coupling_cutoff}
\end{figure}

With these considerations in mind, we are invited to reinterpret the quantization condition for the ground state, Eq. \eqref{eq: QCgroundstate}, to instead consider it as a definition for how the running coupling $g(\Lambda)$ should change as we change the cutoff $\Lambda$ so as to keep expressions for physical observables fixed. We also remark that determining the running in this way produces a multivalued running coupling $g_k(\Lambda)$ depending on the integral factor $k$. We therefore should write Eq. \eqref{eq: QCgroundstate} more appropriately
\begin{align}\label{eq:runningcoupling}
 g_k(\Lambda) \ln \frac{\Lambda_{IR}}{\Lambda} + \Arg \tilde{I}_{ig_k(\Lambda)} \left( \frac{2\Lambda_{IR}}{\Lambda} \right)  + (2k+1) \pi = 0  \ ,
\end{align}
as the equation that in principle \emph{exactly} determines how the coupling runs. Thus Eq. \eqref{eq:runningcoupling} will now be referred to as the running coupling condition. However we will usually just write $g$ although it should be implicitly understood that actually $g = g_k(\Lambda)$.  In Fig. \ref{fig: 1D_ISP_coupling_cutoff} we provide a numerical plot of the running coupling.

We now want to understand what happens to the running coupling as we try to remove the cutoff for some fixed ground state $\Lambda_{IR}$. Hence, we take $\frac{\Lambda_{IR}}{\Lambda} \ll 1$ for the moment and generalize later. In this limit we can approximate
\begin{align} 
\Arg \tilde{I}_{ig} \left( \frac{2\Lambda_{IR}}{\Lambda} \right) = -\Arg \Gamma(1+ig) + \ldots  \ , \qquad \frac{\Lambda_{IR}}{\Lambda} \ll 1 \ ,
\end{align}
and so the running coupling condition at large energy scales becomes
\begin{align}
g \ln \frac{\Lambda_{IR}}{\Lambda} - \Arg \Gamma(1+ig) + (2k+1)\pi = 0 \ , \qquad \frac{\Lambda_{IR}}{\Lambda} \ll 1 \ .
\end{align}
We now show that this has solutions at least for small values of the coupling. Indeed taking $g_k (\Lambda)  \ll 1$ at large cutoff scales, it follows $\frac{\Lambda_{IR}}{\Lambda} \ll 1$ and we can expand as a power series in $g$. With $ \Arg \Gamma(1+ig)  = - \gamma g +\ldots$ we obtain
\begin{align}
(2k+1)\pi  - \left( - \gamma + \ln \frac{\Lambda}{\Lambda_{IR}} \right) g + O(g^2) = 0 \ , \qquad \frac{\Lambda_{IR}}{\Lambda} \ll 1 \ .
\end{align}
Here $\gamma = 0.577721...$ is the Euler-Mascheroni constant. The coupling therefore should run as a function of the cutoff in the following way
\begin{align}\label{eq:runningcouplingfirstorder}
g_k(\Lambda) =  \frac{ (2k+1)\pi}{ - \gamma +  \ln  \frac{\Lambda}{\Lambda_{IR}} } \ , \qquad  \frac{\Lambda_{IR}}{\Lambda} \ll 1 \ , \qquad g \ll 1 \ .
\end{align}
As the cutoff is removed the coupling goes to zero. This implies that the system goes to a non-trivial fixed point $\tilde{c}_0 \rightarrow \frac{1}{4}$ or equivalently $c_0 \rightarrow \frac{1}{4} \frac{\hbar^2}{2m}$ where scale symmetry is recovered. Of course we could also have arrived at this (more easily) from the expression for the ground state in Eq. \eqref{groundstate} which should not be taken to depend on the cutoff. Indeed keeping only the first term in $f(g)$ the expression is
\begin{align}
\Lambda_{IR} = \Lambda e^{- \left(  \frac{ (2k+1) \pi}{g_k(\Lambda)} + \gamma \right)} \ ,
\end{align}
which is just a rewriting of Eq. \eqref{eq:runningcouplingfirstorder}.

It will be one of the major goals of this work to compute how the coupling runs to higher orders in a transseries expansion that includes both perturbative and non-perturbative contributions. Before pursuing this goal however we should go back, and for the sake of completeness, consider what happens to the remaining $n \neq1 $ states within our setup. This is possible at large scales $\Lambda \rightarrow \infty$ where the coupling $g(\Lambda) \to 0$ and things are under control. Define the scale of the $n$'th state to be
\begin{align}
\Lambda_n = \hbar \kappa_n \ , \qquad E_n = - \frac{\Lambda_n^2}{2m} \ , \qquad n \neq 1  \ .
\end{align}
Then the quantization condition in Eq. \ref{eq:QCn} for the $n\neq 1$ states is
\begin{align}
g \ln \frac{\Lambda_n}{\Lambda} + \Arg \tilde{I}_{ig} \left( \frac{2\Lambda_n}{\Lambda} \right) + (2k+n) \pi = 0 \ , \qquad n \neq 1 \ ,
\end{align}
where the running coupling $g = g_k(\Lambda)$ is determined from Eq. \ref{eq:runningcoupling}. We want to know what happens to $\Lambda_n$ with $n \neq 1$ as we try to remove the cutoff. The bound state scales $\Lambda_n$ with $n\neq 1$ can no longer be taken to be fixed and independent of the cutoff $\Lambda$. There are then two possibilities. As we take $\Lambda \rightarrow \infty$ either $\Lambda_n \rightarrow \infty$ or $\Lambda_n$ approaches a finite value. In the former case the bound state energy diverges $E_n \rightarrow - \infty$ and so we simply discard these states. The states with $n<0$ fall into this category as we shall see. Therefore instead consider the latter case where $\Lambda_n$ approaches a finite value as we take $\Lambda \rightarrow \infty$. In this limit we can then take $\frac{\Lambda_n}{\Lambda} \ll 1$ so that $g_k(\Lambda) \rightarrow 0$ and approximate the single-valued argument  term similarly to above to find
\begin{align}
\Lambda_n = \Lambda e^{-\left( \frac{(2k+n)\pi}{g} + \gamma \right)} = \Lambda_{IR} e^{- \frac{(n-1)\pi}{g}} \ .
\end{align}
Since $\Lambda_{IR}$ does not change as $\Lambda \rightarrow \infty$ while $g_k(\Lambda) \rightarrow 0$ the states with $n>1$ all have $\Lambda_n$ approaching zero. Hence, all these state are squeezed out and disappear from the spectrum. Clearly  states with $n<0$ do not have $\Lambda_n$ approaching a finite value and have $\Lambda_n \rightarrow \infty$ as $\Lambda \rightarrow \infty$ so that their energies diverge $E_n \rightarrow - \infty$ and hence should be discarded. 

Moving on we want to find the running coupling to higher orders than what is given in Eq. \ref{eq:runningcouplingfirstorder}. To solve the running coupling condition Eq. \ref{eq:runningcoupling} in terms of a transseries we first write it as a power series in $g$. This is possible since $\Arg \tilde{I}_{ig}$ has a power series in $g$. The gamma function $\Gamma (m+ig+1)$ has no poles in the positive complex plane and has a power series in $g$. The exact running coupling condition Eq. \ref{eq:runningcoupling} is then written in the following way
\begin{align}
 0 =(2k+1)\pi +  \sum_{n=0}^{\infty} a_{2n+1} g^{2n+1} = (2k+1) \pi + a_1 g + a_3 g^3 + a_5 g^5 + \ldots \ ,
\end{align}
where each coefficient is
\begin{align}
a_1 & = \gamma   - \ln \frac{\Lambda}{\Lambda_{IR}} - \left( \frac{\Lambda_{IR}}{\Lambda} \right)^2  + \frac{5}{8} \left( \frac{\Lambda_{IR}}{\Lambda}  \right)^4 -  \frac{23}{54} \left( \frac{\Lambda_{IR}}{\Lambda}  \right)^6  + \dots \ , \\
a_3 & = \frac{1}{6} \psi^{(2)}(1)  + \left( \frac{\Lambda_{IR}}{\Lambda} \right)^2 - \frac{49}{32} \left( \frac{\Lambda_{IR}}{\Lambda} \right)^4 + \frac{3647}{1944} \left( \frac{\Lambda_{IR}}{\Lambda} \right)^6 + \ldots \ , \\
a_5 & = -\frac{1}{120}  \psi^{(4)} (1) - \left( \frac{\Lambda_{IR}}{\Lambda} \right)^2 + \frac{321}{128} \left( \frac{\Lambda_{IR}}{\Lambda} \right)^4 - \frac{320039}{69984} \left( \frac{\Lambda_{IR}}{\Lambda} \right)^6  + \ldots  \ , \\
& \vdots 
\end{align}
Note that the first term in the coefficient $a_1$ has the logarithm term which is non-analytic at $\Lambda \rightarrow \infty$. Clearly we expect the solution to depend on the reciprocal of the non-analytic logarithm term since we have already seen this in a first approximation above. But of course, it will depend on $\frac{\Lambda_{IR}} {\Lambda}$ too to some even power since it only enters the coefficients $a_{2n+1}$ in even powers. Both $\left(\ln \frac{\Lambda}{\Lambda_{IR}} \right)^{-1} \to 0$ and $\frac{\Lambda_{IR}}{\Lambda} \to 0$ as we take $\Lambda \to \infty$. We therefore make the following ansatz for our transseries solution
\begin{align}
\label{eq: bound sector coupling ansatz}
g_k(\Lambda) = \sum_{i=0}^{\infty} \sum_{l=1}^{\infty} c_{2i,l} \left( \frac{\Lambda_{IR}}{\Lambda} \right)^{2i} \left( \ln \frac{\Lambda}{\Lambda_{IR}} \right)^{-l} \ .
\end{align}
As we will now see this ansatz solves the running coupling condition in a unique and simple way order by order to any order we wish. To avoid cumbersome notation we denote $\rho = \left( \ln \Lambda /\Lambda_{IR} \right)^{-1}$, $\xi= \Lambda_{IR}/\Lambda$ and $2k+1=n$. We then plug the ansatz into the running coupling condition and expand as follows
\begin{align}
0 & = \bigg[  (n\pi - c_{0,1} ) \rho + ( \gamma c_{0,1} - c_{0,2} ) \rho^2 +(\gamma c_{0,2} - c_{0,3}) \rho^3 \nonumber \\
&  + \left(\gamma c_{0,3} - c_{0,4} + \frac{1}{6} c_{0,1}^3 \psi^{(2)}(1) \right) \rho^4 + \ldots \bigg]  \nonumber \\
& + \bigg[-c_{2,1} \rho + (-c_{0,1} + \gamma c_{2,1} - c_{2,2} ) \rho^2 + (-c_{0,2} +\gamma c_{2,2} -c_{2,3}) \rho^3   \nonumber \\
&  + \left(c_{0,1}^3 - c_{0,3} + \gamma c_{2,3} - c_{2,4} +\frac{1}{2} c_{0,1}^2 c_{2,1} \psi^{(2)}(1) \right) \rho^4 + \ldots \bigg] \xi^2 \nonumber \\
& + \dots \ .
\end{align}
This equation can only hold provided each coefficient vanishes. From the term $\xi^0$ we can solve recursively for $c_{0,1}$, then $c_{0,2}$ and then $c_{0,3}$ and so on. This can be done linearly to any order we wish.  From the term $\xi^2$ we can solve first for $c_{2,1}$, then for $c_{2,2}$ and then for $c_{2,3}$ and so on. Again this can be done linearly to any order we wish. We find for the $\mathcal{O}(\xi^0)$ coefficients
\begin{align}
\label{eq: bound sector zero coefficients}
c_{0,1} & = n\pi \ , \\
c_{0,2} & = n\pi\gamma \ , \\
c_{0,3} & = n\pi\gamma^2 \ , \\
c_{0,4} & = n\pi\gamma^3 + \frac{1}{6} n^3\pi^3 \psi^{(2)}(1) \ , \\
c_{0,5} & = n\pi\gamma^4 +  \frac{4}{6} \gamma n^3\pi^3 \psi^{(2)}(1)  \, \\
c_{0,6} & = n\pi\gamma^5 + \frac{10}{6} \gamma^2 n^3\pi^3 \psi^{(2)}(1) - \frac{1}{120} n^5\pi^5 \psi^{(4)}(1) \ , \\
c_{0,7} &= n\pi\gamma^6 + \frac{20}{6}\gamma^3 n^3\pi^3 \psi^{(2)}(1) - \frac{6}{120} n^5 \pi^5 \gamma \psi^{(4)}(1) + \frac{1}{12} n^5 \pi^5 \psi^{(2)^2}(1)  \, \\
c_{0,8} &= n \pi \gamma^7 + \frac{35}{6} \gamma^4 n^3 \pi^3 \psi^{(2)}(1) - \frac{21}{120} n^5 \pi^5 \gamma^2 \psi^{(4)}(1) + \frac{7}{12} n^5 \pi^5 \gamma \psi^{(2)^2}(1)\nonumber \ ,\\
&+ \frac{1}{5040} n^7 \pi^7 \psi^{(6)}(1) \ , \\
c_{0,9} &= n \pi \gamma^8 + \frac{46}{6} \gamma^5 n^3 \pi^3 \psi^{(2)}(1) - \frac{56}{120} n^5 \pi^5 \gamma^3 \psi^{(4)}(1) + \frac{28}{12} n^5 \pi^5 \gamma^2 \psi^{(2)^2}(1)\nonumber \ , \\
&+ \frac{8}{5040} n^7 \pi^7 \gamma \psi^{(6)}(1) - \frac{1}{90} n^7 \pi^7 \psi^{(2)}(1) \psi^{(4)}(1) \ .
\end{align}
In appendix \ref{sec: coefficients} we provide $c_{2,l}$ and $c_{4,l}$ coefficients. We have written each coefficient in a very suggestive way in "columns" which allows us to see a pattern and sum up each column:
\begin{align}
&n\pi \rho \sum_{l=0}^\infty (\gamma \rho)^l = n\pi\frac{\rho}{1-\gamma \rho} = \frac{n \pi}{-\gamma + \ln \frac{\Lambda}{\Lambda_{IR}}}  \ ,\\
&n^3 \pi^3 \rho^4 \frac{\psi^{(2)}(1)}{6} \sum_{l=0}^{\infty} \frac{(l+3)!}{3! l!}  (\gamma \rho)^l = \frac{1}{6}\frac{n^3 \pi^3}{\left(-\gamma + \rho^{-1}\right)^4} \psi^{(2)}(1) \ ,\\
&- n^5 \pi^5 \rho^6 \frac{1}{120} \psi^{(4)} \sum_{l=0}^{\infty} \frac{(l+5)!}{5! l!} (\gamma \rho)^l = \frac{-1}{120} \frac{n^5 \pi^5\psi^{(4)}(1)}{\left(-\gamma + \rho^{-1}\right)^6} \ ,\\
&n^5 \pi^5 \rho^7 \frac{\psi^{(2)^2}(1)}{12} \sum_{l=0}^{\infty} \frac{(l+6)!}{6! l!} (\gamma \rho)^l  = \frac{1}{12} \frac{n^5 \pi^5\psi^{(2)^2}(1)}{\left(-\gamma + \rho^{-1}\right)^7} \ .
\end{align}
We have checked that the above pattern also holds for $c_{0,10}$, $c_{0,11}$, $c_{0,12}$ and $c_{0,13}$. Continuing in a similar way yields the final result
\begin{align}
\label{eq: bound sector total running coupling}
g_k\left(\Lambda\right) &= \bigg[\frac{(2k+1) \pi}{-\gamma + \ln \frac{\Lambda}{\Lambda_{IR}} + \left(\frac{\Lambda_{IR}}{\Lambda}\right)^2} + \frac{1}{6} \frac{(2k+1)^3 \pi^3}{\left(-\gamma + \ln \frac{\Lambda}{\Lambda_{IR}} + \left(\frac{\Lambda_{IR}}{\Lambda}\right)^2\right)^4} \psi^{(2)}(1)\nonumber\\
&- \frac{1}{120} \frac{(2k+1)^5 \pi^5}{\left(-\gamma + \ln \frac{\Lambda}{\Lambda_{IR}} + \left(\frac{\Lambda_{IR}}{\Lambda}\right)^2\right)^6} \psi^{(4)}(1) + \frac{1}{12} \frac{(2k+1)^5 \pi^5}{\left(-\gamma + \ln \frac{\Lambda}{\Lambda_{IR}} + \left(\frac{\Lambda_{IR}}{\Lambda}\right)^2\right)^7} \psi^{(2)^2}(1)\nonumber\\
&+ \frac{1}{5040} \frac{(2k+1)^7 \pi^7}{\left(-\gamma + \ln \frac{\Lambda}{\Lambda_{IR}} + \left(\frac{\Lambda_{IR}}{\Lambda}\right)^2\right)^8} \psi^{(6)}(1) - \frac{1}{90}\frac{(2k+1)^7 \pi^7}{\left(-\gamma + \ln \frac{\Lambda}{\Lambda_{IR}} + \left(\frac{\Lambda_{IR}}{\Lambda}\right)^2\right)^9} \psi^{(2)}(1) \psi^{(4)}(1) + \ldots\bigg]\nonumber\\
&+ \bigg[\frac{(2k+1)^3 \pi^3}{\left(-\gamma + \ln \frac{\Lambda}{\Lambda_{IR}} + \left(\frac{\Lambda_{IR}}{\Lambda}\right)^2\right)^4} - \frac{(2k+1)^5 \pi^5}{\left(-\gamma + \ln \frac{\Lambda}{\Lambda_{IR}} + \left(\frac{\Lambda_{IR}}{\Lambda}\right)^2\right)^6}\nonumber\\
&+ \frac{(2k+1)^5 \pi^5}{\left(-\gamma + \ln \frac{\Lambda}{\Lambda_{IR}} + \left(\frac{\Lambda_{IR}}{\Lambda}\right)^2\right)^7} \psi^{(2)}(1) + \ldots\bigg] \left(\frac{\Lambda_{IR}}{\Lambda}\right)^2\nonumber\\
&+ \bigg[\frac{5}{8} \frac{(2k+1) \pi}{\left(-\gamma + \ln \frac{\Lambda}{\Lambda_{IR}} + \left(\frac{\Lambda_{IR}}{\Lambda}\right)^2\right)^2} - \frac{49}{32} \frac{(2k+1)^3 \pi^3}{\left(-\gamma + \ln \frac{\Lambda}{\Lambda_{IR}} + \left(\frac{\Lambda_{IR}}{\Lambda}\right)^2\right)^4}\nonumber\\
&+ \frac{5}{12} \frac{(2k+1)^3 \pi^3}{\left(-\gamma + \ln \frac{\Lambda}{\Lambda_{IR}} + \left(\frac{\Lambda_{IR}}{\Lambda}\right)^2\right)^5} \psi^{(2)}(1) + \ldots \bigg]\left(\frac{\Lambda_{IR}}{\Lambda}\right)^4\\
&\vdots\nonumber
\end{align}
Looking at the running coupling we now learn a number of things. Coming from our original ansatz Eq. \eqref{eq: bound sector coupling ansatz} we have been able to guess and sum up various parts of the solution so we could write it in the form above. The result suggests that the final structure of the running coupling is of the form
\begin{align}
g_k(\Lambda) = \sum_{p=0}\sum_{q=1} c^{(k)}_{p,q} \frac{1}{\left(-\gamma + \ln \frac{\Lambda}{\Lambda_{IR}} + \left(\frac{\Lambda_{IR}}{\Lambda}\right)^2\right)^q} \left(\frac{\Lambda_{IR}}{\Lambda}\right)^p \ .
\end{align}

\subsubsection{The Beta Function}

Instead of attempting directly an approximate solution to $g(\Lambda)$ from the running coupling condition it is customary to define a new function which determines the scaling of the coupling through a first order differential equation. This is the beta function.  This is easy to arrive at since we have already solved for the ground state $\Lambda_{IR} = \Lambda f(g(\Lambda))$ where $f(g)$ is given in terms of a transseries in Eq. \eqref{eq: bound sector f result}.  Acting with the operator $\frac{d}{d\Lambda} $ on $\Lambda_{IR} = \Lambda f(g(\Lambda))$ we find
\begin{align}
 0 = f(g) + \frac{\partial f(g)}{\partial g} \Lambda \frac{d g}{d \Lambda} \ , 
\end{align}
or if we define the beta function of the coupling we have 
\begin{align}
\label{eq: 1D_ISP_bound_beta_def}
\beta(g) \equiv \Lambda \frac{dg}{d\Lambda} = - f(g) \left( \frac{\partial f(g)}{\partial g} \right)^{-1} = \frac{-1}{\partial_g \ln f(g)} \ .
\end{align}
The beta function is of fundamental importance and tells us how the coupling changes as we vary the cutoff. Equation \eqref{eq: f ansatz1} provides a simple way of obtaining \textit{exact} results order by order in the non-perturbative exponential. Without loss of generality we take $k=0$. Then
\begin{align}\label{eq: frikadelle}
\beta(g) &= \frac{-\frac{1}{\pi} g^2}{1+ \frac{g^2}{\pi} \left(\frac{1}{g} \Arg \Gamma(1+ig)\right)'}\nonumber\\
&+\frac{\frac{2}{\pi^2} g^2}{(1+g^2)^2} \frac{\pi + (\pi - g)g^2 + (1+g^2)g^2 \left(\frac{1}{g} \Arg \Gamma(1+ig)\right)'}{\left(1+ \frac{g^2}{\pi} \left(\frac{1}{g} \Arg \Gamma(1+ig)\right)'\right)^2} e^{2\frac{\Arg \Gamma(1+ig)}{g}} e^{-2\frac{\pi}{g}}\nonumber\\
&+ \frac{\frac{1}{\pi^3} g^2}{(4+g^2)^2\left(1+\frac{g^2}{\pi} \left(\frac{1}{g} \Arg \Gamma(1+ig)\right)'\right)^3} \bigg[\frac{6 \pi^2 (1+g^2)^3}{4+g^2} \left(1+\frac{g^2}{\pi} \left(\frac{1}{g} \Arg \Gamma(1+ig)\right)'\right)^2\nonumber\\
&-\frac{g^3 \pi (1+g^2) (-49-11g^2+2g^4)}{(4+g^2)^2}\left(1+\frac{g^2}{\pi} \left(\frac{1}{g} \Arg \Gamma(1+ig)\right)'\right) \nonumber\\
&-4 g^6 \bigg] e^{4 \frac{\Arg \Gamma(1+ig)}{g}} e^{-4 \frac{\pi}{g}} + \ldots \ .
\end{align}
Here $'$ denotes derivative with respect to $g$. Note that the perturbative part is completely determined by the leading order in $\Lambda_{IR}/\Lambda$ in Eq. \eqref{eq: 1D_ISP_bound_state_poly}. Explicitly evaluating the derivatives yields the following transseries expansion
\begin{align}
\label{eq: bound sector beta totale}
\beta(g) &= \left[\frac{-1}{\pi} g^2 - \frac{\psi^{(2)}(1)}{3 \pi^2} g^5 + \frac{\psi^{(4)}(1)}{30 \pi^2} g^7 - \frac{\psi^{(2)^2}(1)}{9 \pi^3} g^8 - \frac{\psi^{(6)}(1)}{840 \pi^2} g^9 + \ldots\right]\nonumber\\
&+ \left[\frac{2}{\pi} g^2 - \frac{2(3+\psi^{(2)}(1))}{3 \pi} g^4 + \frac{2(-3 + \psi^{(2)}(1))}{3 \pi} g^5\ldots\right]e^{-2\left( \frac{\pi}{g}+\gamma\right)}\nonumber\\
&+ \left[\frac{3}{2 \pi} g^2 - \frac{\left(15+8\psi^{(2)}(1)\right)}{8 \pi}g^4 + \frac{\left(49+8\psi^{(2)}(1)\right)}{16 \pi^2} g^5 + \ldots\right] e^{-4 \left(\frac{\pi}{g} + \gamma\right)}\nonumber\\
&+ \left[ \frac{37}{18\pi} g^2 - \frac{2677+1332\psi^{(2)}(1)}{648\pi}g^4 + \frac{7529+1332\psi^{(2)}(1)}{1944\pi^2}g^5 + \ldots       \right] e^{-6 \left(\frac{\pi}{g} + \gamma\right)}\nonumber\\
&+ \left[\frac{3113}{864 \pi} g^2 - \frac{11 (115039 + 54336 \psi^{(2)}(1)}{124416 \pi} g^4 + \frac{3002027 + 597696 \psi^{(2)}(1)}{497664 \pi^2} g^5 + \ldots\right]e^{-8 \left(\frac{\pi}{g} + \gamma\right)} \ .
\end{align}
The main results arrived at in the bound state sector consitute Eq.s \eqref{eq: QCgroundstate}, \eqref{eq: bound sector f result}, \eqref{eq: bound sector total running coupling} and \eqref{eq: bound sector beta totale}. In Eq. \eqref{eq: QCgroundstate} we present the exact quantum condition for the $1/x^2$ potential on the positive axis with the Dirichlet boundary condition $\psi(\epsilon)=0$ and in Eq.s \eqref{eq: bound sector f result}, \eqref{eq: bound sector total running coupling} and \eqref{eq: bound sector beta totale} we compute the renormalization of the model to very high order. Some results presented are even exact.

\subsection{Scattering Sector $E>0$}

We now move to scattering states $E>0$ and write the general solution Eq. \ref{eq:sol1} as a linear combination of the two Hankel functions (See App. \ref{App: bessel} for a review)
\begin{align}
\psi_k(x) & = (kx)^{\frac{1}{2}} \left( C_- H_{ig}^{(1)}(kx) +C_+ H_{ig}^{(2)} (kx) \right)  \ , \qquad C_{\mp} = \frac{A \mp i B}{2}  \ .
\end{align}
The asymptotic forms of the Hankel functions are
\begin{align}
H_{ig}^{(1)} (kx) & = \sqrt{ \frac{2}{\pi kx} } e^{i \left( kx - \frac{i  \pi g }{2} - \frac{\pi}{4} \right)} G_{ig}^{(1)} (k x) \ , \\
H_{ig}^{(2)} (kx) & = \sqrt{ \frac{2}{\pi kx} } e^{- i \left( kx - \frac{i  \pi g }{2} - \frac{\pi}{4} \right)}G_{ig}^{(2)} (k x) \ ,
\end{align}
with
\begin{align}
G_{ig}^{(1)} (k x) & = \sum_{n=0}^{\infty} i^n \frac{a_n(ig)}{(kx)^n}  = 1 - i \frac{1+4g^2}{8kx} + O \left( \frac{1}{(kx)^2}\right) \ , \\
G_{ig}^{(2)} (k x) &=\sum_{n=0}^{\infty} (-i)^n \frac{a_n(ig)}{(kx)^n} = 1  + i \frac{1+4g^2}{8kx} + O \left( \frac{1}{(kx)^2}\right) \ , \\
a_n(ig) & =  \frac{(4(ig)^2-1^2)(4(ig)^2-3^2)\cdots(4(ig)^2-(2n-1)^2)}{n! 8^n} \ , \qquad a_0(ig) = 1 \ .
\end{align}
Then, the asymptotic form of the scattering solution is
\begin{align}
\psi_k(x) &= \sqrt{ \frac{2}{\pi k} } C_+  e^{i \left(  \frac{ig\pi}{2} + \frac{\pi}{4}  \right)} \left( e^{- i kx} - i  \frac{C_-}{C_+} e^{\pi g  } e^{ i  kx   } \right)   \left[ 1+ O\left( \frac{1}{kx} \right)\right] \ .
\end{align}
The first term represents an ingoing left-moving wave while the second term represents an outgoing (reflected) right-moving wave. The phase shift $\delta$ is defined asymptotically via
\begin{align}
e^{i 2\delta} \equiv   i  \frac{C_-}{C_+} e^{\pi g  }  \ .
\end{align}
The ratio of the coefficients $C_{\mp}$ is found by enforcing the boundary condition
\begin{align}
\psi_k (\epsilon) = 0 \qquad \Rightarrow \qquad \frac{C_-}{C_+} = - \frac{H_{ig}^{(2)} (k\epsilon) }{H_{ig}^{(1)} (k\epsilon)}  \ ,
\end{align}
and so the phase shift is given by the following condition
\begin{align}
e^{i2\delta} &= -  i  \frac{H_{ig}^{(2)} (k\epsilon) }{H_{ig}^{(1)} (k\epsilon)} e^{\pi g}  \ .
\end{align} 
As for the bound state case, we rewrite $k$ and $\epsilon$ in terms of momentum scales $p = \hbar k$ and $\Lambda =2\hbar/ \epsilon$. Then the phase shift is given by
\begin{align}
e^{i2\delta} &= -  i  \frac{H_{ig}^{(2)} \left(\frac{2p}{\Lambda} \right) }{H_{ig}^{(1)} \left(\frac{2p}{\Lambda} \right)} e^{\pi g}  \ .
\end{align} 
Although we have written an expression for the phase shift in a neat and compact form we want to write an equation which is manifestly real. Using that the argument of the Hankel functions are real and the identities provided in Eq.s \eqref{eq: hankel relaton1} and \eqref{eq: hankel relaton2} we obtain
\begin{align}
\delta = - \frac{\pi}{4} - \Arg H^{(1)}_{ig} \left(\frac{2p}{\Lambda}\right) \ .
\end{align}
Now take $\pi/4$ to the other side and instead write
\begin{align}
\tan\left(\delta + \frac{\pi}{4} \right) = - \frac{ \text{Im} ~ H^{(1)}_{ig}}{\text{Re} ~ H^{(1)}_{ig}} \ .
\end{align}
Using the definition of $H^{(1)}_{ig}$ in terms of the ordinary Bessel functions of the first and second kind we may write
\begin{align}
H^{(1)}_{ig} &= \left[1+\coth \pi g\right] J_{ig} - \frac{1}{\sinh \pi g} J_{-ig}\\
&= |J_{ig}| \left(\left[1+\coth \pi g\right] e^{i \Arg J_{ig}} - \frac{1}{\sinh \pi g} e^{-i \Arg J_{ig}}\right) \ .
\end{align}
The expression is of the form
\begin{align}
z = A e^{i\theta} - B e^{-i \theta} = (A-B) \cos \theta + i (A+B) \sin \theta \ .
\end{align}
Then
\begin{align}
\frac{\text{Im} ~ z}{\text{Re} ~ z} = \frac{A+B}{A-B} \tan \theta  \ .
\end{align}
With $A=1+\coth \pi g$ and $B=1/\sinh \pi g$ we get
\begin{align} \label{eq: phase shift equation}
\tan\left(\delta + \frac{\pi}{4}\right)  + \coth \left( \frac{\pi}{2}g \right) \tan\left[\Arg J_{ig}\left(\frac{2p}{\Lambda}\right)\right] =0 \ .
\end{align}
This result is the exact version of the approximate result presented in Eq. 3.19 in \cite{Gupta_1993}. This expression for the phase shift is what we will use in the following when we renormalize the theory. 

\subsubsection{Renormalization from the Scattering Sector}
We now treat Eq. \eqref{eq: phase shift equation} similarly to the quantization condition in the bound state sector and reinterpret it as a condition for the running coupling. First note, however, that here in the scattering sector the momentum $p$ appears in the above equation and so the phase shift must be a function $\delta(p)$. Now take the phase shift $\delta(p)$ to be a physical observable\footnote{The phaseshift is not directly a physical observable, however it is for example related to time delay.} that should be independent of the cutoff and imagine that we have measured it to be $\delta (p_0)$ at some particular momentum $p = p_0$. Then we instead take the coupling to carry all of the cutoff dependence $ g(\Lambda)$. The coupling should then vary with the cutoff such that the phase shift stays constant.  The equation for the phase shift becomes an equation for the running coupling and should then more appropriately be written as
\begin{align}
\tan\left(\delta(p_0) + \frac{\pi}{4}\right)  + \coth \left( \frac{\pi}{2} g (\Lambda)  \right)  \tan\left[\Arg J_{ig(\Lambda)}\left(\frac{2p_0}{\Lambda}\right)\right] =0 \ .
\end{align}
From here on we do not explicitly write the $\Lambda$ dependence of the coupling. In a similar way as we decomposed $\Arg I_{ig}$ we can decompose $\Arg J_{ig}$ into
\begin{align}
\Arg J_{ig}\left(\frac{2p}{\Lambda}\right) &= g \ln \frac{p}{\Lambda} - \Arg \Gamma(1+ig) + \Arg \tilde{\eta}_{ig}\left(\frac{p}{\Lambda}\right),\\
\tilde{\eta}_{ig}\left(\frac{p}{\Lambda}\right) &= 1 + \sum_{m=1}^{\infty} \frac{(-1)^m}{m!}\left(\prod_{n=0}^{m-1}\frac{1}{1+ig+n}\right) \left(\frac{p}{\Lambda}\right)^{2m} \ .
\end{align}
Furthermore we will work with 
\begin{align}
K = \frac{1}{2} \tan\left(\delta + \frac{\pi}{4}\right) \ ,
\end{align}
and write the phase condition as
\begin{align}
	\label{eq: frikadelle}
K + \frac{1}{2}\coth \left( \frac{\pi}{2} g \right) \tan\left[g \ln \frac{p}{\Lambda} - \Arg \Gamma(1+ig) + \Arg \tilde{\eta}_{ig}\left( \frac{p}{\Lambda} \right) \right] = 0 \ .
\end{align}
Following the bound state sector Sec. \ref{sec: bound sector} we expect a transseries dependence of the coupling on the cutoff $\Lambda$. Looking at the argument of the tangens function in Eq. \eqref{eq: frikadelle} we arrive at the ansatz
\begin{align}
g_n = \sum_{k=0}^{\infty} \sum_{l=1}^{\infty} c_{2 k, l} \left(\frac{p}{\Lambda}\right)^{2k} \left(\ln \frac{\Lambda}{p}\right)^{-l}, \quad c_{0,1} = n \pi, \quad n \in \mathbb{N} \ .
\end{align}
The choice $c_{0,1}=n \pi$ allows us to fix each coefficient linearly order by order. The factor of $n$ is the freedom one has from the periodicity of $\tan x$. Plugging the ansatz into the phase condition leads to (c.f. Eq. \eqref{eq: bound sector zero coefficients})
\begin{align}
c_{0,1} &= n \pi \ , \\
c_{0,2} &= n \pi \left(\gamma + K \pi\right) \ , \\
c_{0,3} &= n \pi \left(\gamma + K \pi\right)^2 \ , \\
c_{0,4} &= n \pi \left(\gamma + K \pi\right)^3 - \frac{1}{12}n^3 \pi^3 \left(K \pi^3 + 4 K^3 \pi^3 - 2 \psi^{(2)}(1)\right) \ , \\
c_{0,5} &= n \pi \left(\gamma + K \pi \right)^4 - \frac{4}{12} n^3 \pi^3 \left(\gamma + K \pi\right)  \left(K \pi^3 + 4 K^3 \pi^3 - 2 \psi^{(2)}(1)\right) \ , \\
c_{0,6} &= n \pi \left(\gamma + K \pi \right)^5 - \frac{10}{12} n^3 \pi^3 \left(\gamma + K \pi\right)^2  \left(K \pi^3 + 4 K^3 \pi^3 - 2 \psi^{(2)}(1)\right)\nonumber\\
&+ \frac{1}{120}n^5 \pi^5 \left(K \pi^5 + 10K^3 \pi^5 + 24 K^5 \pi^5 - \psi^{(4)}(1)\right) \ ,\\
c_{0,7} &= n \pi \left(\gamma + K \pi\right)^6 - \frac{20}{12} n^3 \pi^3 \left(\gamma + K \pi\right)^3 \left(K \pi^3 + 4 K^3 \pi^3 - 2 \psi^{(2)}(1)\right)\nonumber\\
&+ \frac{6}{120} n^5 \pi^5 \left(\gamma + K \pi \right) \left(K \pi^5 + 10K^3 \pi^5 + 24 K^5 \pi^5 - \psi^{(4)}(1)\right)\nonumber\\
&+ \frac{1}{48} n^5 \pi^5 \left(K \pi^3 + 4 K^3 \pi^3 - 2 \psi^{(2)}(1)\right)^2 \ , \\
\vdots\nonumber
\end{align}
Coefficients $c_{2, l}$ and $c_{4,l}$ can be found in App. \ref{sec: coefficients}. As in the bound state sector Sec. \ref{sec: bound sector} we can sum up the various columns order by order in $p/\Lambda$ to find (c.f. Eq. \eqref{eq: bound sector total running coupling})
\begin{align} \label{eq: scattering coupling}
g_n (\Lambda) &= \bigg[\frac{n \pi}{\ln \frac{\Lambda}{p} - \left(\gamma + K\pi + \left(\frac{p}{\Lambda}\right)^2\right)} - \frac{1}{12} \frac{n^3 \pi^3 \left(K \pi^3 + 4 K^3 \pi^3 - 2 \psi^{(2)}(1)\right)}{\left(\ln \frac{\Lambda}{p} - \left(\gamma + K\pi + \left(\frac{p}{\Lambda}\right)^2\right)\right)^4} +\nonumber\\
&+ \frac{1}{120} \frac{n^5 \pi^5 \left(K \pi^5 + 10 K^3 \pi^5 + 24 K^5 \pi^5 - \psi^{(4)}(1)\right)}{\left(\ln \frac{\Lambda}{p} - \left(\gamma + K\pi + \left(\frac{p}{\Lambda}\right)^2\right)\right)^6} \ldots\bigg]\nonumber\\
&+ \bigg[-\frac{n^3\pi^3}{\left(\ln \frac{\Lambda}{p} - \left(\gamma + K\pi + \left(\frac{p}{\Lambda}\right)^2\right)\right)^4} + \ldots \bigg]\left(\frac{p}{\Lambda}\right)^2\nonumber\\
&+ \bigg[\frac{5}{8} \frac{n \pi}{\left(\ln \frac{\Lambda}{p} - \left(\gamma + K\pi + \left(\frac{p}{\Lambda}\right)^2\right)\right)^2} + \ldots \bigg] \left(\frac{p}{\Lambda}\right)^4  \ .
\end{align}
We compute the beta function directly from this coupling for $n=1$.
\begin{align}
\beta = \Lambda \dv{g}{\Lambda} = -\frac{1}{\ln^2 \frac{\Lambda}{p}} \pdv{g}{1/\ln (\Lambda/p)} - \frac{p}{\Lambda} \pdv{g}{p/\Lambda}  \ .
\end{align}
We now replace $p/\Lambda$ with $g$ by directly inverting the expression for the coupling in Eq. \eqref{eq: scattering coupling} via a similar ansatz as Eq. \eqref{eq: f ansatz1} but for $p/\Lambda$ to find (c.f. Eq. \eqref{eq: bound sector beta totale})
\begin{align}
\label{eq: beta from scattering}
\beta &= \left[\frac{-1}{\pi}g^2 + \frac{K \pi^3 + 4 K^3 \pi^3 - 2 \psi^{(2)}(1)}{6 \pi^2} g^5 + \ldots\right]\nonumber\\
&+ \bigg[-\frac{2}{\pi} g^2 - \frac{-6 + K \pi^3 + 4 K^3 \pi^3 - 2 \psi^{(2)}(1)}{3 \pi} g^4\nonumber\\
&+ \frac{6 + K \pi^3 + 4 K^3 \pi^3 - 2 \psi^{(2)}(1)}{3 \pi^2}g^5 + \ldots \bigg]e^{-2 K \pi} e^{-2\left( \frac{\pi}{g} + \gamma \right)}\nonumber\\
&+ \left[ \frac{3 }{2\pi} g^2 + \ldots\right] e^{-4 K \pi} e^{-4\left( \frac{\pi}{g} + \gamma \right)}\nonumber\\
&\vdots
\end{align}
The overall structure of the beta function is the same as the one found in the bound state sector Eq. \eqref{eq: bound sector beta totale}. There are simply some sign differences as well as the presence of the scattering phase $\delta(p)$ in the scattering sector. Consequently, it appears as if the two couplings, computed in each sector, runs differently. To make this statement more explicit, one can consider expanding one coupling in another. This is done by replacing every $p/\Lambda$ in Eq. \eqref{eq: scattering coupling} with $p/\Lambda_{IR} \times f(g_B)$ where $g_B$ is the coupling computed in the bound sector with $k=0$. We then find
\begin{align}
\label{eq: coupling sector expansion}
g_S &= g_B + \left(K - \frac{1}{\pi} \ln \frac{\Lambda_{IR}}{p}\right) g_B^2 + \left(K - \frac{1}{\pi} \ln \frac{\Lambda_{IR}}{p}\right)^2g_B^3\nonumber\\
&+ \left(\left(K - \frac{1}{\pi} \ln \frac{\Lambda_{IR}}{p}\right)^3 - \frac{1}{12} K \left(1+4K^2\right)\right)g_B^4\nonumber\\
&+  \left(K - \frac{1}{\pi} \ln \frac{\Lambda_{IR}}{p}\right) \left(\left(K - \frac{1}{\pi} \ln \frac{\Lambda_{IR}}{p}\right)^3 - \frac{K \pi^3 + 4 K^3 \pi^3}{3 \pi}\right) g_B^5 + \ldots \nonumber\\
&+ \left[\frac{1}{\pi}\left(1 + \left(\frac{p}{\Lambda_{IR}}\right)^2\right)g_B^2 + \frac{2}{\pi} \left(1+\left(\frac{p}{\Lambda_{IR}}\right)^2\right)\left(K - \frac{1}{\pi} \ln \frac{\Lambda_{IR}}{p} \right)g_B^3 + \ldots \right] e^{-2 \frac{\pi}{g_B}}\nonumber\\
&\vdots
\end{align}
where $g_S$ is the coupling computed in the scattering sector with $n=1$. Note that the two couplings are related non-perturbatively. However, the first perturbative coefficient of the two beta functions remains unchanged. The consequence is that the running of the scattering and bound sector couplings meet in the vicinity of the UV fixed point where scale symmetry is recovered. Therefore, by going to the fixed point we then obtain an equation solely in terms of physical observables
\begin{align}
\ln \frac{\Lambda_{IR}}{p_0} = \pi K \Leftrightarrow \tan \delta_0 = \frac{\ln \frac{p}{\Lambda_{IR}} + \frac{\pi}{2}}{\ln \frac{p}{\Lambda_{IR}} - \frac{\pi}{2}} \ ,
\end{align}
which agrees with the result found in \cite{Griffiths:1DISP}. This can be rewritten in terms of two momenta
\begin{align}
\tan \delta'(p_1) = \tan \delta'(p_0) - \frac{2}{\pi}\ln \frac{p_1}{p_0} \ ,
\end{align}
which agrees with the result found in \cite{Gupta_1993}. In the low energy scattering regime $p\ll \Lambda_{IR}$ we note that $\delta_0 \to \pi/4$ or equivalently $K \to \infty$. To check the consistency of these equations we can employ the usual analytical continuation between the bound state and scattering sectors $p \to i \Lambda_{IR}$. Consequently
\begin{align}
K(p \to i \Lambda_{IR}) = \frac{1}{\pi} \ln \frac{\Lambda_{IR}}{(p \to i \Lambda_{IR})} = -\frac{i}{2} \ .
\end{align} 
Inserting this into Eq. \eqref{eq: coupling sector expansion} it is now a simple task to check that indeed the scattering coupling maps directly into the bound sector coupling.

\subsection{The $S$-Matrix}
\noindent
The S-Matrix is simply
\begin{align}
S = e^{i 2 \delta} = -i \frac{H_{ig}^{(2)}\left(\frac{2p}{\Lambda}\right)}{H_{ig}^{(1)}\left(\frac{2p}{\Lambda}\right)} e^{\pi g} \ .
\end{align}
We identify bound states as poles in the S-Matrix represented in the complex plane of $k$ or equivalently $p$. Hence we want to rewrite
\begin{align}
H_{ig}^{(1)} \left(2 \frac{(p\to i\Lambda_{IR})}{\Lambda}\right) = 0 \ .
\end{align}
The replacement $p \to i \Lambda_{IR}$ should be understood as an analytic continuation and not as a direct replacement as we require both $p^2>0$ and $\Lambda_{IR}^2 > 0$. Then
\begin{align}
H_{ig}^{(1)}\left(2i \frac{\Lambda_{IR}}{\Lambda}\right) &= \frac{e^{\pi g} J_{ig}\left(2i \frac{\Lambda_{IR}}{\Lambda}\right) - J_{-ig}\left(2i \frac{\Lambda_{IR}}{\Lambda}\right)}{\sinh{\pi g}}\\
&= \frac{e^{\pi g} e^{-\frac{\pi}{2} g} I_{ig}\left(2 \frac{\Lambda_{IR}}{\Lambda}\right) - e^{\frac{\pi}{2} g} I_{-ig}\left(2 \frac{\Lambda_{IR}}{\Lambda}\right)}{\sinh{\pi g}}\\
&= 0 \ .
\end{align}
This gives the condition
\begin{align}
I_{ig}\left(2 \frac{\Lambda_{IR}}{\Lambda}\right) = I_{-ig}\left(2 \frac{\Lambda_{IR}}{\Lambda}\right) \ .
\end{align}
Using, that the two Bessel functions have the same modulo but opposite phase we get
\begin{equation}
e^{i 2\arg I_{ig}\left(2 \frac{\Lambda_{IR}}{\Lambda}\right)} = 1 \ , 
\end{equation}
which leads to
\begin{align}
\arg I_{ig} \left(\frac{2 \Lambda_{IR}}{\Lambda}\right) + n \pi = 0 \ .
\end{align}
Exactly what we found in the bound state sector Eq. \eqref{eq: bound state argI condition}.


\newpage

\section{Outlook}\label{conclusions}
\noindent
Computing and interpreting renormalization group flows remain a novel task in quantum theories. We began this work by perturbatively studying the divergences in scale invariant quantum mechanical systems with multiple couplings. While this provides a simple illustration of the core difficulties in dealing with classically scale invariant mechanical systems in quantum theories, at the core of our work is, the unravel of a rich non-perturbative instanton-like structure in the RG flow of the inverse square potential. Particularly, we provided exactly the full infinite perturbative part of the beta function as well as two complete non-perturbative orders in Eq. \eqref{eq: frikadelle} where higher orders may be obtained trivially.  The coupling differs depending on whether it is renormalized in the bound state sector or in scattering sector. However, following the flow to the fixed point of the theory, the couplings become identical and yield an unambiguous physical prediction.

This work opens new possible windows for exploring RG flows with non-perturbative features in a setting where explicit analytical control may be gained. In particular, it has  laid the ground for our further and more complete investigations of multiple coupling systems and how their perturbative and non-perturbative dynamics interplay on each other.

\acknowledgments
We thank Chat-GPT for minor discussions and latex help. 

\newpage
\appendix

\section{Conformal Group and Symmetry}\label{App: conformal symmetry}

The models considered in this work all have time translation and scale symmetry. The former follows from the explicit time independence while the latter follows from the observation that there are no parameters with dimension. They are in fact also invariant under special conformal transformations. This happens to be the case since the Hamiltonian is no more than quadratic in ${\bf p}$. For  scale vs conformal invariance see for instance the review \cite{Nakayama:2013is}. We refer to these three symmetry transformations as conformal transformations and they form together the conformal group $\text{PSL}(2,\mathbb{R}) = \text{SL}(2,\mathbb{R})/\mathbb{Z}_2$ which is the projective special linear group. To see how all of this pans out consider the following elegant approach.

At the abstract level acting on the one dimensional temporal line conformal transformations are defined as the real Möbius transformations
\begin{align}\label{ConTransT}
t' = \frac{a t + b}{c t + d}  \ , \qquad ad - bc =1 \ .
\end{align}
On the trajectory the conformal transformations are realized as
\begin{align}\label{ConTransX}
{\bf x}'(t') = \frac{1}{c t + d} {\bf x}(t) \ ,
\end{align}
implying 
\begin{align}
\dot{\bf x}'(t') = - c {\bf x}(t) + (ct+d) \dot{\bf x}(t) \ .
\end{align}
There are four parameters $a,b,c,d$ in a conformal transformation but due to the determinant condition $ad-bc = 1$ there are only three independent parameters. We choose to write these as 
\begin{align}
&\text{Time translation:} &t'&  = t + t_0 \ ,& {\bf x}'(t')& = {\bf x}(t) \\
&\text{Scale transformation:} &t'& =\lambda^{2} t \ , &{\bf x}'(t')& = \lambda {\bf x}(t) \\
&\text{Special conformal transformation:} &t'& =\frac{1}{1-\alpha t} t \ , &{\bf x}'(t')& = \frac{1}{1-\alpha t} {\bf x}(t)  \ .
\end{align}
The scaling exponent of $t$ is twice the scaling exponent of ${\bf x}$ in agreement with the respective dimensions being $\dim t = \text{L}^2$ and $\dim {\bf x} = \text{L}$. Time translations are arrived at by choosing $c=0,a=d=1,b=t_0$, scale transformations (dilatations) are arrived at by choosing $b=c=0,a=d^{-1} = \lambda$ and special conformal transformations are arrived at by choosing $a=d=1, b=0, c=- \alpha$. Note that we always have $\lambda^2>0$ in the scale transformation so the discrete time reversal transformation $t'=-t$ is not part of the conformal transformations. A perhaps better way to consider a special conformal transformation is to write it as 
\begin{align}
\frac{1}{t'} = \frac{1}{t} - \alpha \ .
\end{align}
Therefore a special conformal transformation consists first of an inversion, then a translation and then an inversion again.  Acting with conformal transformations on the time coordinate is equivalent to multiplying $2 \times 2$ matrices of the form
\begin{align}
\mathbb{M} & = 
\begin{pmatrix}
a & b \\
c & d
\end{pmatrix} \ , \qquad ad -bc =1 \ , \qquad \mathbb{M} \in \text{SL}(2,\mathbb{R}) \ .
\end{align}
The special linear group SL$(2,\mathbb{R})$ is connected and non-compact. Note that a conformal transformation by $\mathbb{M}\in \text{SL}(2,\mathbb{R})$ is the same as a conformal transformation by $-\mathbb{M}\in \text{SL}(2,\mathbb{R})$ when acting on the time coordinate. Therefore flipping the sign of $\mathbb{M}$ does not matter and the two conformal transformations induced by $\mathbb{M}$ and $-\mathbb{M}$ respectively are the same when acting on the time coordinate. The conformal group must therefore be the projective special linear group PSL$(2,\mathbb{R})$=SL$(2,\mathbb{R})/\mathbb{Z}_2$. In this group theory language we can write a time translation, a small dilatation ($\lambda = 1 + \frac{\epsilon}{2} + \ldots$) and a special conformal transformation as
\begin{align}
\mathbb{M}_{\mathbb{H}}& = 
\begin{pmatrix}
1 & t_0 \\
0 & 1
\end{pmatrix} = 
\mathbb{I} - t_0 \mathbb{H} \ , &\mathbb{H}& = 
\begin{pmatrix}
0 & -1 \\
0 & 0
\end{pmatrix} \ , \\
\mathbb{M}_{\mathbb{D}}& = 
\begin{pmatrix}
\lambda & 0 \\
0 & \lambda^{-1}
\end{pmatrix} = 
\mathbb{I} - \epsilon \mathbb{D} + \ldots  \ , &\mathbb{D}& = 
\frac{1}{2} \begin{pmatrix}
-1 & 0 \\
0 & 1
\end{pmatrix} \ , \\
\mathbb{M}_{\mathbb{K}}& = 
\begin{pmatrix}
1 & 0 \\
-\alpha & 1
\end{pmatrix} = 
\mathbb{I} - \alpha \mathbb{K} \ , &\mathbb{K}& = 
\begin{pmatrix}
0 & 0 \\
1 & 0
\end{pmatrix} \ .
\end{align}
The three generators $\mathbb{H},\mathbb{D},\mathbb{K}$ are all traceless as they should be since elements of the conformal group have unit determinant. They also satisfy the Lie algebra of the conformal group 
\begin{align}\label{eqLieAlgebraAbstract}
\left[ \mathbb{D},\mathbb{H}\right] =- \mathbb{H} \ , \qquad \left[ \mathbb{D}, \mathbb{K} \right] = \mathbb{K} \ , \qquad \left[ \mathbb{K}, \mathbb{H} \right] =  - 2\mathbb{D} \ .
\end{align}
It is sometimes convenient to write the generators in another basis 
\begin{align}
\mathbb{M}^1 = -\frac{1}{2} (\mathbb{H}+\mathbb{K}) \ , \qquad \mathbb{M}^2 = - \frac{1}{2} (\mathbb{H}-\mathbb{K}) \ , \qquad \mathbb{M}^3 = - \mathbb{D} \ ,
\end{align}
in which the Lie algebra of the conformal group is
\begin{align}
\left[\mathbb{M}^i , \mathbb{M}^j  \right] =   \epsilon^{ijk} \eta_{kl} \mathbb{M}^l \ , \qquad \eta_{ij} = \text{diag}(-1,1,1) \ .
\end{align} 
Written in this way we see that the Lie algebra of the conformal group is isomorphic to the Lie algebra of SO$(1,2)$. The group SO$(1,2)$ contains two components and the component containing the identity is the group denoted SO$^+(1,2)$. The conformal group PSL$(2,\mathbb{R})$ is isomorphic to SO$^+(1,2)$. 

We are now ready to prove that the system with Hamiltonian given by Eq. \ref{theory} is conformal invariant. Namely if  we consider the (Lagrangian) action then it is invariant under conformal transformations as given in Eq.'s \ref{ConTransT} and \ref{ConTransX} up to a boundary term
\begin{align}
\int_{t_a'}^{t_b'}  L({\bf x}', \dot{\bf x}' )  \dd t' = \int_{t_a}^{t_b} L({\bf x}, \dot{\bf x})  \dd t + \int_{t_a}^{t_b} \frac{d}{dt} \left( -\frac{m}{2}  \frac{c{\bf x}^2}{c t + d}  \right)  \dd t \ .
\end{align}
where the Lagrangian is
\begin{align}
L({\bf x}, \dot{\bf x}) = \frac{1}{2} m  \dot{\bf x}^2 - V(r) \ , \qquad V(r) = - \frac{c}{r^2}- k \frac{\delta(r)}{r} \ .
\end{align}

\section{Bessel Functions and Useful Relations}\label{App: bessel}
We state in this section useful definitions and properties of Bessel functions used throughout the main text taken from \cite{Table_of_Integrals_Series_Products}. The two linearly independent solutions to Bessels equation
\begin{align}
	\label{eq: bessels equation}
	z^2 \dv[2]{Z}{z} + z \dv{Z}{z} + (z^2-\nu^2)Z = 0,
\end{align}
are known as Bessel functions of the first and second kind respectively given by the series representation
\begin{align}
	J_{\nu}(z) &= \sum_{m=0}^\infty \frac{(-1)^m}{m! \Gamma(\nu+m+1)} \left(\frac{z}{2}\right)^{\nu+2m},\\
	Y_{\nu}(z) &= \frac{J_{\nu}(z) \cos{\pi \nu} - J_{-\nu}(z)}{\sin{\pi \nu}} .
\end{align}
This definition of $Y_\nu$ requires $\nu$ not to be an integer. The Hankel functions of the first and second kind are another pair of linearly independent solutions given by repsectively
\begin{align}
	H^{(1)}_\nu(z) &= J_\nu(z) + i Y_\nu(z),\\
	H^{(2)}_\nu(z) &= J_\nu(z) - i Y_\nu(z).
\end{align}
They satisfy (Eq.s. $4.484$ in \cite{Table_of_Integrals_Series_Products})
\begin{align} 
	\label{eq: hankel relaton1}
	H^{(1)}_{-\nu}(z) &= e^{i \pi \nu} H_{\nu}^{(1)}(z),\\
	\label{eq: hankel relaton2}
	H^{(2)}_{-\nu}(z) &= e^{-i \pi \nu} H_\nu^{(2)}(z).
\end{align}

The two linearly independent solutions to the modified Bessel equation
\begin{align}
	\label{eq: modified bessels equation}
	z^2 \dv[2]{Z}{z} + z \dv{Z}{z} - (z^2+\nu^2)Z = 0,
\end{align}
are the modified Bessel functions of the first and second kind respectively given by the series representation
\begin{align}
	I_\nu(z) &= \sum_{m=0}^\infty \frac{1}{m! \Gamma(\nu+m+1)} \left(\frac{z}{2}\right)^{\nu+2m},\\
	K_\nu(z) &= \frac{\pi}{2} \frac{I_{-\nu}(z) - I_\nu(z)}{\sin \pi \nu},
\end{align}
where again the latter definition requires $\nu$ to be a non-integer.

\section{Fourier Transforms}\label{Fourier}

In this appendix we calculate the Fourier transforms needed to find the matrix elements $\langle {\bf p}' |V| {\bf p} \rangle $. In $d$ spatial dimensions define the Fourier transform of a function $f({\bf x})$ as
\begin{align}
\tilde{f} ({\bf q})  &= \mathcal{F}\left[f(\vb{x})\right] = \int d^d x f({\bf x}) e^{- i {\bf x}\cdot {\bf q}} \ , \\
f(\vb{x}) &= \mathcal{F}^{-1}\left[\tilde{f}(\vb{q})\right] = \int \frac{\dd^d q}{(2\pi)^d} ~ \tilde{f}(\vb{q}) e^{i \vb{q}\cdot \vb{x}} \ .
\end{align}
A couple of observations that are useful now follows. If the function $f$ is even $f(- {\bf x}) = f({\bf x})$ then the Fourier transform is even $\tilde{f} (- {\bf k}) = \tilde{f} ({\bf q}) $. And if the function $f$ is odd $f(- {\bf x}) = - f({\bf x})$ then the Fourier transform is odd $\tilde{f} (- {\bf q}) =- \tilde{f} ({\bf q}) $. 

For all the matrix elements we insert a complete set of position basis kets and arrive at a Fourier transform which then has to be evaluated. The wave function of a momentum eigenstate in $d$ spatial dimensions is normalized as
\begin{align}
\langle {\bf x} | {\bf p} \rangle= \frac{1}{(2\pi \hbar)^{d/2}} e^{i \frac{ {\bf x} \cdot {\bf p} }{\hbar} } \qquad \text{and} \qquad \langle  {\bf p}  |  {\bf x} \rangle= \frac{1}{(2\pi \hbar)^{d/2}} e^{ - i \frac{ {\bf x} \cdot {\bf p} }{\hbar} } \ .
\end{align}
For ease of notation in $d>1$ we also define ${\bf q} = \frac{{\bf p'} - {\bf p} }{\hbar}$ and $q = |{\bf q}| $
while in $d=1$ we define $q = \frac{p'-p}{\hbar}$.

\subsection{$d=1$}
For the first matrix element we write
\begin{align}
  (2\pi\hbar) \langle p' | \frac{-c}{r^2} | p \rangle & = -c \int_{\mathbb{R}} \dd x \frac{ 1 }{x^2} e^{- i x q } =  c  \int_{\mathbb{R}} \dd x \frac{d}{dx} \left( \frac{1}{x}\right) e^{-ixq} \nonumber \\
  & =  -c  ( - i q) \int_{\mathbb{R}} \dd x \frac{1}{x} e^{-ixq}  =   -c  ( - i q) \mathcal{F}\left[ x^{-1} \right]  \ .
\end{align}
where in the last equality we have performed a partial integration. The boundary terms vanish. We have now rewritten it as the Fourier transform of $x^{-1}$. To find it first consider taking its derivative 
\begin{align}
 \frac{i}{2\pi} \frac{d}{dq} \mathcal{F}\left[ x^{-1} \right]  =\frac{1}{2\pi} \int_{\mathbb{R}} \dd x e^{-iqx} =  \delta(q) \ ,
\end{align}
and therefore the Fourier transform must be the Heaviside step function
\begin{align}
  F\left( x^{-1} \right)  = - i  2\pi   \theta(q) + i \pi \ .
\end{align}
The constant $i \pi$ has been chosen so that the Fourier transform $\tilde{f} (q) = \mathcal{F}\left[ x^{-1} \right] $ is an odd function in $q$ which must be the case since $x^{-1}$ is an odd function. Therefore
\begin{align}
\mathcal{F}\left[ x^{-1} \right]  = - i \pi \sgn (q) \ .
\end{align}
Putting it all together we find for the matrix element
\begin{align}
  (2\pi\hbar) \langle p' | \frac{-c}{r^2} | p \rangle & = c \pi |q| = c \pi \frac{|p'-p|}{\hbar} \ .
\end{align} 
For the second matrix element we write 
\begin{align}
( 2\pi \hbar ) \langle p' | \frac{-k \delta (|x|)}{|x|} | p \rangle & = k \int_{\mathbb{R}} \dd x \sgn(x) \delta'(x) e^{- i x q } \ .
\end{align}
This integral is divergent and we need to regularize it. We choose to regulate the sign function and take it to be 
\begin{align}
\sgn(x) &= \lim_{\epsilon \to 0} \frac{ x}{{\sqrt{x^2 + \epsilon^2}}} \ ,
\end{align}
and evaluate the Fourier integral by partial integration
\begin{align}
\int_{-\infty}^{\infty} \dd x \frac{ x}{{\sqrt{x^2 + \epsilon^2}}}  \delta'(x) e^{- i x q } = - \int_{-\infty}^{\infty} \dd x \delta(x)  \frac{\epsilon^2 - i q x (x^2+\epsilon^2)}{(x^2 + \epsilon^2)^{3/2}} e^{-i q x}  =  -\frac{1}{\epsilon} \ .
\end{align}
The boundary terms vanish because the Dirac delta function vanishes at positive/negative infinity. As we try to remove the cutoff $\epsilon \to 0$ the integral diverges. We can also switch to a large momentum cutoff $\Lambda  = 2\hbar / \epsilon$ for which the divergence is then linear
\begin{align}
\int_{-\infty}^{\infty} \dd x \frac{ x}{{\sqrt{x^2 + \epsilon^2}}}  \delta'(x) e^{- i x q } = -\frac{1}{2\hbar} \Lambda \ .
\end{align}
The matrix element is 
\begin{align}
( 2\pi \hbar ) \langle p' | \frac{-k \delta (|x|)}{|x|} | p \rangle & = -\frac{k }{2\hbar} \Lambda \ .
\end{align}
For the parity odd matrix element we use $\delta(x) = - x \delta'(x)$ and write 
\begin{align}
(2\pi \hbar)\langle p' | \frac{-k' \delta(x)}{x} | p \rangle & = k' \int_{\mathbb{R}} \dd x \delta'(x) e^{-i x	q } = i k' \frac{p'-p}{\hbar} \ .
\end{align}

\subsection{$d=2$}
\noindent
For the first matrix element we write in polar coordinates
\begin{align}
(2\pi \hbar)^2 \langle {\bf p'} | \frac{- c}{r^2} | {\bf p} \rangle & = - c \int_{0}^{\infty} \dd r \frac{1}{r} \int_{0}^{2\pi} \dd \theta  e^{-i r q  \cos \theta} = -2\pi c \int_{0}^{\infty} \dd (qr) \frac{J_0(qr)}{qr} \ ,
\end{align}
where $J_0(qr)$ is the Bessel function of the first kind. This integral diverges as $r \to 0$. The integrand has small and large argument expansions
\begin{align}
\frac{J_0(qr)}{q r} & = \frac{1}{q r} - \frac{qr}{4}+ O((q r)^3)  \\
\frac{J_0(qr)}{qr} & = \frac{1}{q} \sqrt{\frac{2}{\pi q r}} \left( \cos \left( q r -\frac{\pi}{4} \right) + O((qr)^{-1}) \right) \ .
\end{align}
If we cut off the integral, for fixed $q$, at small values of $qr$ and at large values of $qr$ we write
\begin{align}
\text{small $q r$} : \qquad &  \int_{q \epsilon} \dd (qr) \frac{J_0(qr)}{qr}   \sim  - \ln q \epsilon + \frac{1}{8} (q\epsilon)^2 + O((q\epsilon)^4) \\
\text{large $q r$}: \qquad &\int^{qL} \dd (qr)  \frac{J_0(qr)}{qr} \sim \frac{a_1(qL)}{q} \frac{1}{(qL)^{1/2}} + \frac{a_2(qL)}{q} \frac{1}{(qL)^{3/2}} + O((qL)^{-5/2}) \ ,
\end{align}
where $\epsilon$ is a small distance regulator and $L$ is a large distance regulator. The coefficients $a_1(qL)$ and $a_2(qL)$ are oscillatory functions of $qL$. Clearly one can remove the large distance cutoff $L \to \infty$ and there is no divergence. However we cannot remove the small distance cutoff $\epsilon \to 0$ as there is a logarithmic divergence.

We see that the integral can be written as 
\begin{align}
\int_{q \epsilon}^{\infty} \dd (qr) \frac{J_0(qr)}{qr} &= - \ln q \epsilon + \frac{1}{8} (q\epsilon)^2 + O((q\epsilon)^4) + \text{finite terms} \ ,
\end{align}
where the finite terms are independent of $q$ and $\epsilon$. In particular as we try to remove the cutoff $\epsilon \to 0$ the all higher order polynomial terms in $q\epsilon$ tend to zero and the dominant term is the logarithm. If we instead switch to a large momentum cutoff $\Lambda =2\hbar / \epsilon$ the integral is
\begin{align}
\int_{\frac{2q \hbar }{\Lambda}}^{\infty} \dd (qr) \frac{J_0(qr)}{qr} &= - \ln \frac{q\hbar}{\Lambda}+ \frac{1}{2} \left(\frac{q\hbar}{\Lambda} \right)^2 + O\left(  \Lambda^{-4} \right)  + \text{finite terms}\\
& = \ln \frac{\Lambda}{|{\bf p'}-{\bf p}|}+ \frac{1}{2} \left(\frac{|{\bf p'}-{\bf p}|}{\Lambda} \right)^2 + O\left(  \Lambda^{-4} \right)  + \text{finite terms} \ ,
\end{align}
where we have moved a constant factor of $\ln 2$ into the finite part. Therefore the matrix element is
\begin{align}
(2\pi \hbar)^2 \langle {\bf p'} | \frac{- c}{r^2} | {\bf p} \rangle & = - 2\pi c \ln \frac{\Lambda}{|{\bf p'}-{\bf p}|} - \pi c \left(\frac{|{\bf p'}-{\bf p}|}{\Lambda} \right)^2 + O\left(  \Lambda^{-4} \right)  + \text{finite terms} \ .
\end{align}
For the second matrix element we simply write
\begin{align}
(2\pi\hbar)^2 \langle {\bf p'} | \frac{-k \delta(r)}{r} | {\bf p} \rangle & = (2\pi\hbar)^2 \langle {\bf p'} | - 2\pi k \delta^{(2)}({\bf x}) | {\bf p} \rangle  = - 2 \pi k  \int_{\mathbb{R}^2} \dd ^2 x \  \delta^2({\bf x}) e^{-i {\bf x} \cdot {\bf q}}  = -  2\pi k \ .
\end{align}

\subsection{ $ d\geq 3$ }
\noindent
In $d\geq 3$ spherical coordinates consist of a radial coordinate $r$ and $d-1$ angles $\phi_i$, $i=1,\ldots,d-1$. The angles $\phi_i$, $i=1,\ldots,d-2$ range over the interval $[0,\pi]$ while the last angle $\phi_{d-1}$ range over the interval $[0,2\pi[$. The volume element is 
\begin{align}
d^d x = r^{d-1} \sin^{d-2}\phi_1 \ldots \sin \phi_{d-2} ~ \dd r \dd \phi_1 \ldots \dd \phi_{d-2} \dd \phi_{d-1} = r^{d-1} ~ \dd r \dd \Omega_d \ .
\end{align}
Consider now
\begin{align}
(2\pi \hbar)^d \bra{\vb{p}'} \frac{-c}{r^2} \ket{\vb{p}} = -c \int_0^\infty \dd r r^{d-3} \int_{S^{d-1}} \dd \Omega_d e^{-i \vb{q} \cdot \vb{x}} \ .
\end{align}
We can choose to orient the spherical coordinate system so that the angle between ${\bf x}$ and ${\bf q}$ is $\phi_{d-2}$. Then we may write
\begin{align}
(2\pi \hbar)^d \bra{\vb{p}'} \frac{-c}{r^2} \ket{\vb{p}} = - \frac{c \Omega_d}{2} \int_0^\infty \dd r r^{d-3} \int_{-1}^{1} \dd \cos \phi_{d-2} e^{-iqr \cos{\phi_{d-2}}}  \ .
\end{align}
We can do the final angular integration and obtain
\begin{align}
(2\pi \hbar)^d \bra{\vb{p}'} \frac{-c}{r^2} \ket{\vb{p}} &= -\frac{c \Omega_d}{q^{d-2}} \int_0^\infty \dd (qr) (qr)^{d-4} \sin(qr) \nonumber \\
& = -\frac{c \Omega_d}{q^{d-2}} \int_0^\infty \dd \mu ~  \mu^{d-4} \sin \mu \ .
\end{align}
The integral on the right-hand-side does not depend on $q$ and so the Fourier transform of $1/r^2$ must be proportional to $1/q^{d-2}$. Evaluating this integral to get the proportionality constant however is slightly tricky. We will make use of two different well known representations of the delta function in $d$ spatial dimensions written in the following way
\begin{align}
\delta^{(d)}(\vb{x}) =  \int_{\mathbb{R}^d} \frac{\dd^d q}{(2\pi)^d}  ~ e^{i \vb{q} \cdot \vb{x}}  \qquad \delta^{(d)}(\vb{x}) = \frac{1}{(d-2) \Omega_d} \nabla^2 \frac{-1}{r^{d-2}} \ .
\end{align}
Now rewrite the first expression as
\begin{align}
\delta^{(d)}(\vb{x}) =  - \mathbb{\nabla}^2 \int_{\mathbb{R}^d} \frac{\dd^d q}{(2\pi)^d}  \frac{1}{q^2}  e^{i \vb{q} \cdot \vb{x}} = - \mathbb{\nabla}^2 \left( \frac{1}{(2\pi)^d} \frac{\Omega_d}{r} \int_{0}^{\infty} \dd q ~ q^{d-4} \sin{qr} \right)  \ ,
\end{align}
where we have switched to spherical coordinates and performed the angular integrations as done previously. Again we can define $\mu = qr$  so that the integral becomes independent of $r$ and then write
\begin{align}
\delta^{(d)}(\vb{x}) =  \frac{\Omega_d}{(2\pi)^d} \left(  \mathbb{\nabla}^2  \frac{ - 1}{r^{d-2}} \right) \int_{0}^{\infty} \dd \mu  ~ \mu^{d-4} \sin \mu  = \frac{(d-2) \Omega_d^2}{(2\pi)^d} \delta^{(d)} ({\bf x}) \int_{0}^{\infty} \dd \mu  ~ \mu^{d-4} \sin \mu   \ ,
\end{align}
or equivalently
\begin{align}
\left( 1 -  \frac{(d-2) \Omega_d^2}{(2\pi)^d} \int_{0}^{\infty} \dd \mu ~ \mu^{d-4} \sin \mu  \right) \delta^{(d)} ({\bf x})= 0  \ .
\end{align}
At the point ${\bf x} = 0 $ this is only possible provided 
\begin{align}
\int_{0}^{\infty}\dd \mu ~ \mu^{d-4} \sin \mu = \frac{(2\pi)^d}{(d-2) \Omega_d^2} \ .
\end{align}
With the integral in hand the matrix element is
\begin{align}
(2\pi \hbar)^d \bra{\vb{p}'} \frac{-c}{r^2} \ket{\vb{p}} = - c \frac{(2\pi)^d}{(d-2) \Omega_d} \frac{\hbar^{d-2}}{|{\bf p}'-{\bf p}|^{d-2}} \ .
\end{align}
For the second matrix element we write in spherical coordinates in $d\geq 3$ spatial dimensions
\begin{align}
(2\pi\hbar)^d \langle {\bf p'} |   \frac{-k \delta(r)}{r}   | {\bf p} \rangle & = -k \int_0^\infty \dd r \delta(r) r^{d-2} \int_{S^{d-1}} \dd \Omega_d e^{-i \vb{q} \cdot \vb{x}} \nonumber \\
& = - \frac{k\Omega_d}{q^{d-2}} \int_{0}^{\infty} \dd (qr) \delta(qr) (qr)^{d-3}\sin(qr) \nonumber \\
& = 0 \ .
\end{align}

\section{Coefficients}\label{sec: coefficients}
\subsection{Bound State Sector}
\noindent
We display here the next two orders in $\Lambda_{IR}/\Lambda$ following Eq. \eqref{eq: bound sector zero coefficients}
\begin{align}
\label{eq: bound sector square coefficients}
c_{2,1} &= 0 \ , \\
c_{2,2} &= - n \pi \ , \\
c_{2,3} &= -2 \gamma n \pi \ , \\
c_{2,4} &= - 3 \gamma^2 n \pi + n^3 \pi^3 \ , \\
c_{2,5} &= - 4 \gamma^3 n \pi + 4 \gamma n^3 \pi^3 - \frac{4}{6} n^3 \pi^3 \psi^{(2)}(1) \ , \\
c_{2,6} &= -5 \gamma^4 n \pi + 10 \gamma^2 n^3 \pi^3 - \frac{20}{6} \gamma n^3 \pi^3 \psi^{(2)}(1) - n^5 \pi^5 \ , \\
c_{2,7} &= -6 \gamma^5 n \pi + 20 \gamma^3 n^3 \pi^3 - \frac{60}{6} \gamma^2 n^3 \pi^3 \psi^{(2)}(1) - 6 \gamma n^5 \pi^5\nonumber\\
&+n^5 \pi^5 \psi^{(2)}(1)+\frac{1}{20}n^5 \pi^5 \psi^{(4)}(1) \ .
\end{align}
Notice that the term multiplying $\psi^{(2)}(1)$ begins at $c_{2,5}$ whereas in the above it began at $c_{0,4}$. This holds for all the other terms as well. At next order we find
\begin{align}
\label{eq: bound sector quadruple coefficients}
c_{4,1} &= 0 \ , \\
c_{4,2} &= \frac{5}{8} n \pi \ , \\
c_{4,3} &= n \pi  + \frac{10}{8} \gamma n \pi \ , \\
c_{4,4} &= 3 \gamma n \pi + \frac{15}{8} \gamma^2 n \pi - \frac{49}{32} n^3 \pi^3 \ , \\
c_{4,5} &= 6 \gamma^2 n \pi + \frac{20}{8} \gamma^3 n \pi - \frac{196}{32} \gamma n^3 \pi^3 - 4 n^3 \pi^3 + \frac{5}{12}n^3 \pi^3 \psi^{(2)}(1) \ , \\
c_{4,6} &= 10 \gamma^3 n \pi + \frac{25}{8}\gamma^4 n \pi - \frac{490}{32} \gamma^2 n^3 \pi^3 - 20 \gamma n^3 \pi^3 + \frac{25}{12} \gamma n^3 \pi^3 \psi^{(2)}(1)\nonumber\\ 
&+ \frac{5}{3} n^3 \pi^3 \psi^{(2)}(1) \ .
\end{align}

\subsection{Scattering Sector}
\begin{align}
c_{2,1} &= 0 \ , \\
c_{2,2} &= n \pi \ , \\
c_{2,3} &= 2 n \pi \left(\gamma + K \pi \right) \ , \\
c_{2,4} &= 3 n \pi \left(\gamma + K \pi \right)^2 - n^3 \pi^3 \ , \\
c_{2,5} &= 4 n \pi \left(\gamma + K \pi \right)^3 - 4 n^3 \pi^3 \left(\gamma + K \pi \right)\nonumber\\
&- \frac{4}{12} n^3 \pi^3 \left(K \pi^3 + 4 K^3 \pi^3 - 2 \psi^{(2)}(1)\right) \ , \\
c_{2,6} &= 5 n \pi \left(\gamma + K \pi\right)^4 - 10 n^3 \pi^3 \left(\gamma + K \pi \right)^2\nonumber\\
&- \frac{20}{12} n^3 \pi^3 \left(\gamma + K \pi \right) \left(K \pi^3 + 4 K^3 \pi^3 - 2 \psi^{(2)}(1)\right) + n^5 \pi^5 \ , \\
c_{2,7} &= 6 n \pi \left(\gamma + K \pi \right)^5 - 20 n^3 \pi^3 \left(\gamma + K \pi \right)^3\nonumber\\
&- \frac{60}{12}n^3 \pi^3 \left(\gamma + K \pi \right)^2 \left(K \pi^3 + 4 K^3 \pi^3 - 2 \psi^{(2)}(1)\right) + 6 n^5 \pi^5 \left(\gamma + K \pi \right)\nonumber\\
&+ \frac{6}{120} n^5 \pi^5 \left(K \pi^5 + 10 K^3 \pi^5 +24 K^5 \pi^5 - \psi^{(4)}(1)\right)\nonumber\\
&+ \frac{1}{2} n^5 \pi^5 \left(K\pi^3 + 4 K^3 \pi^3 - 2 \psi^{(2)}(1) \right) \ .
\end{align}

\begin{align}
c_{4,1} &= 0 \ , \\
c_{4,2} &= \frac{5}{8} n \pi \ , \\
c_{4,3} &= \frac{10}{8} n \pi \left(\gamma + K \pi\right) + n \pi \ , \\
c_{4,4} &= \frac{15}{8} n \pi \left(\gamma + K \pi \right)^2 + 3 n \pi \left(\gamma + K \pi \right) - \frac{49}{32} n^3 \pi^3 \ , \\
c_{4,5} &= \frac{20}{8} n \pi \left(\gamma + K \pi \right)^3 + 6 n \pi \left(\gamma + K \pi \right)^2 - \frac{196}{32} n^3 \pi^3 \left(\gamma + K \pi \right) - 4 n^3 \pi^3\nonumber\\
&- \frac{5}{24} n^3 \pi^3 \left(K \pi^3 + 4 K^3 \pi^3 - 2 \psi^{(2)}(1)\right) \ , \\
c_{4,6} &= \frac{25}{8} n \pi \left(\gamma + K \pi \right)^4 + 10 n \pi \left(\gamma + K \pi \right)^3 - \frac{490}{32} n^3 \pi^3 \left(\gamma + K \pi \right)^2 - 20 n^3 \pi^3 \left(\gamma + K \pi \right)\nonumber\\
&- \frac{25}{24} n^3 \pi^3 \left(\gamma + K \pi \right) \left(K \pi^3 + 4 K^3 \pi^3 - 2 \psi^{(2)}(1)\right) - \frac{20}{24}n^3 \pi^3 \left(K \pi^3 + 4 K^3 \pi^3 -2 \psi^{(2)}(1)\right)\nonumber\\
&+ \frac{321}{128} n^5 \pi^5  \ .
\end{align}

\newpage
\bibliography{QM}

@article{Wilson:1974mb,
    author = "Wilson, Kenneth G.",
    title = "{The Renormalization Group: Critical Phenomena and the Kondo Problem}",
    reportNumber = "CLNS-296",
    doi = "10.1103/RevModPhys.47.773",
    journal = "Rev. Mod. Phys.",
    volume = "47",
    pages = "773",
    year = "1975"
}

@article{Wilson:1973jj,
    author = "Wilson, K. G. and Kogut, John B.",
    title = "{The Renormalization group and the epsilon expansion}",
    doi = "10.1016/0370-1573(74)90023-4",
    journal = "Phys. Rept.",
    volume = "12",
    pages = "75--199",
    year = "1974"
}

@article{Camblong:2000qn,
    author = "Camblong, Horacio E. and Epele, Luis N. and Fanchiotti, Huner and Garcia Canal, Carlos A.",
    title = "{Dimensional transmutation and dimensional regularization in quantum mechanics. 1. General theory}",
    eprint = "hep-th/0003255",
    archivePrefix = "arXiv",
    doi = "10.1006/aphy.2000.6092",
    journal = "Annals Phys.",
    volume = "287",
    pages = "14--56",
    year = "2001"
}

@article{Camblong:2000ax,
    author = "Camblong, Horacio E. and Epele, Luis N. and Fanchiotti, Huner and Garcia Canal, Carlos A.",
    title = "{Dimensional transmutation and dimensional regularization in quantum mechanics. 2. Rotational invariance}",
    eprint = "hep-th/0003267",
    archivePrefix = "arXiv",
    doi = "10.1006/aphy.2000.6093",
    journal = "Annals Phys.",
    volume = "287",
    pages = "57--100",
    year = "2001"
}

@article{Holstein,
   title={Anomalies in quantum mechanics: The 1/r2 potential},
   volume={70},
   ISSN={1943-2909},
   url={http://dx.doi.org/10.1119/1.1456071},
   DOI={10.1119/1.1456071},
   number={5},
   journal={American Journal of Physics},
   publisher={American Association of Physics Teachers (AAPT)},
   author={Coon, Sidney A. and Holstein, Barry R.},
   year={2002},
   month=may, pages={513–519} }

@article{Griffiths:1DISP,
    author  = {David J. Griffiths and Andrew M. Essin},
    title   = {Quantum mechanics of the $1/x^2$ potential},
    journal = {American Journal of Physics},
    volume  = {74},
    pages   = {109},
    year    = {2006},
    doi     = {10.1119/1.2165248}}

@article{Gupta_1993,
   title={Renormalization in quantum mechanics},
   volume={48},
   ISSN={0556-2821},
   url={http://dx.doi.org/10.1103/PhysRevD.48.5940},
   DOI={10.1103/physrevd.48.5940},
   number={12},
   journal={Physical Review D},
   publisher={American Physical Society (APS)},
   author={Gupta, K. S. and Rajeev, S. G.},
   year={1993},
   month=dec, pages={5940–5945} }

@article{Camblong:2001zt,
    author = "Camblong, Horacio E. and Epele, Luis N. and Fanchiotti, Huner and Garcia Canal, Carlos A.",
    title = "{Quantum anomaly in molecular physics}",
    eprint = "hep-th/0106144",
    archivePrefix = "arXiv",
    doi = "10.1103/PhysRevLett.87.220402",
    journal = "Phys. Rev. Lett.",
    volume = "87",
    pages = "220402",
    year = "2001"
}

@article{Nakayama:2013is,
    author = "Nakayama, Yu",
    title = "{Scale invariance vs conformal invariance}",
    eprint = "1302.0884",
    archivePrefix = "arXiv",
    primaryClass = "hep-th",
    reportNumber = "CALT-68-2910",
    doi = "10.1016/j.physrep.2014.12.003",
    journal = "Phys. Rept.",
    volume = "569",
    pages = "1--93",
    year = "2015"
}

@book{Table_of_Integrals_Series_Products,
  author       = {I.S. Gradshteyn, I.M. Ryzhik},
  title        = {{Table of Integrals, Series and Products}},
  year         = 1980,
  publisher    = {Academic Press},
  version      = 7}

@misc{Luty,
    author      = {Luty},
    title       = {{Physics 851 notes - Renormalization}},
    howpublished = {\url{https://www.physics.umd.edu/courses/Phys851/Luty/notes/renorm.pdf}},
    year        = {2007}
}

@misc{Mcgreevy,
    author      = {John McGreevy},
    title       = {{Physics 215B, Quantum Field Theory, Winter 2018}},
    howpublished = {\url{https://mcgreevy.physics.ucsd.edu/w18/}},
    year        = {2018}
}

@article{FermiTeller1947,
  title = {The Capture of Negative Mesotrons in Matter},
  author = {Fermi, E. and Teller, E.},
  journal = {Phys. Rev.},
  volume = {72},
  issue = {5},
  pages = {399--408},
  numpages = {0},
  year = {1947},
  month = {Sep},
  publisher = {American Physical Society},
  doi = {10.1103/PhysRev.72.399},
  url = {https://link.aps.org/doi/10.1103/PhysRev.72.399}
}

@article{JeanLevy1967,
  title = {Electron Capture by Polar Molecules},
  author = {L\'evy-Leblond, Jean-Marc},
  journal = {Phys. Rev.},
  volume = {153},
  issue = {1},
  pages = {1--4},
  numpages = {0},
  year = {1967},
  month = {Jan},
  publisher = {American Physical Society},
  doi = {10.1103/PhysRev.153.1},
  url = {https://link.aps.org/doi/10.1103/PhysRev.153.1}
}

@book{landau2013quantum,
  title={Quantum mechanics: non-relativistic theory},
  author={Landau, Lev Davidovich and Lifshitz, Evgenii Mikhailovich},
  volume={3},
  year={2013},
  publisher={Elsevier}
}

@article{TurnerJE1966,
author = {Turner, J. E.},
issn = {0031-899X},
journal = {Physical review},
number = {1},
pages = {21-26},
title = {Electron Capture by Rotational Excitation of Polar Molecules},
volume = {141},
year = {1966},
}

@article{JACKIWR1972Iss,
author = {JACKIW, R},
address = {WOODBURY},
copyright = {Copyright 2016 Elsevier B.V., All rights reserved.},
issn = {0031-9228},
journal = {Phys. Today 25: No. 1, 23-7(Jan 1972)},
number = {1},
organization = {Massachusetts Inst. of Tech., Cambridge},
pages = {23-27},
publisher = {AIP Publishing},
title = {Introducing scale symmetry},
volume = {25},
year = {1972},
}

@incollection{DeAlfaroV.2006Ciif,
author = {De Alfaro, V. and Fubini, S. and Furlan, G. and Bleuler, Konrad and Reetz, Axel and Petry, Herbert Rainer},
address = {Berlin, Heidelberg},
booktitle = {Differential Geometrical Methods in Mathematical Physics II},
copyright = {Springer-Verlag 1978},
isbn = {9783540089353},
issn = {0075-8434},
pages = {275-280},
publisher = {Springer Berlin Heidelberg},
series = {Lecture Notes in Mathematics},
title = {Conformal invariance in field theory},
year = {2006},
}

@article{CASEKM1950Sp,
author = {CASE, KM},
address = {COLLEGE PK},
copyright = {Copyright 2007 Elsevier B.V., All rights reserved.},
issn = {0031-899X},
journal = {Physical review},
number = {5},
pages = {797-806},
publisher = {AMERICAN PHYSICAL SOC},
title = {Singular potentials},
volume = {80},
year = {1950},
}

@book{MottMasseyAtomicCollisions,
  title={The Theory of Atomic Collisions},
  author={Mott, N. F. and Massey H. S. W.},
  volume={2},
  year={1949},
  publisher={Clarendon Press, Oxford}
}

@article{fox1966wkb,
  title={WKB Treatment of Bound States in an Electric-Dipole Potential},
  author={Fox, Kenneth and Turner, JE},
  journal={American Journal of Physics},
  volume={34},
  number={7},
  pages={606--610},
  year={1966},
  publisher={AIP Publishing}
}

@article{fox1966variational,
  title={Variational Calculation for Bound States in an Electric-Dipole Field},
  author={Fox, Kenneth and Turner, JE},
  journal={The Journal of Chemical Physics},
  volume={45},
  number={4},
  pages={1142--1144},
  year={1966},
  publisher={American Institute of Physics}
}

@article{turner1968ground,
  title={Ground-state energy eigenvalues and eigenfunctions for an electron in an electric-dipole field},
  author={Turner, James Edward and Anderson, VE and Fox, Kenneth},
  journal={Physical Review},
  volume={174},
  number={1},
  pages={81},
  year={1968},
  publisher={APS}
}

@article{altshuler1957theory,
  title={Theory of low-energy electron scattering by polar molecules},
  author={Altshuler, Saul},
  journal={Physical Review},
  volume={107},
  number={1},
  pages={114},
  year={1957},
  publisher={APS}
}

@article{Holten_2018,
   title={Anomalous Breaking of Scale Invariance in a Two-Dimensional Fermi Gas},
   volume={121},
   ISSN={1079-7114},
   url={http://dx.doi.org/10.1103/PhysRevLett.121.120401},
   DOI={10.1103/physrevlett.121.120401},
   number={12},
   journal={Physical Review Letters},
   publisher={American Physical Society (APS)},
   author={Holten, M. and Bayha, L. and Klein, A. C. and Murthy, P. A. and Preiss, P. M. and Jochim, S.},
   year={2018}
}

@article{Sundaram_2024,
   title={Duality between the quantum inverted harmonic oscillator and inverse square potentials},
   volume={26},
   ISSN={1367-2630},
   url={http://dx.doi.org/10.1088/1367-2630/ad3a91},
   DOI={10.1088/1367-2630/ad3a91},
   number={5},
   journal={New Journal of Physics},
   publisher={IOP Publishing},
   author={Sundaram, Sriram and Burgess, C P and O’Dell, D H J},
   year={2024},
   month=may, pages={053023} }

@article{Sundaram_2021,
   title={Fall-to-the-centre as a 
	    
	    PT
	    
	   symmetry breaking transition},
   volume={2038},
   ISSN={1742-6596},
   url={http://dx.doi.org/10.1088/1742-6596/2038/1/012024},
   DOI={10.1088/1742-6596/2038/1/012024},
   number={1},
   journal={Journal of Physics: Conference Series},
   publisher={IOP Publishing},
   author={Sundaram, Sriram and Burgess, C P and O’Dell, Duncan H J},
   year={2021},
   month=oct, pages={012024} }

@article{frank1971singular,
  title={Singular potentials},
  author={Frank, William M and Land, David J and Spector, Richard M},
  journal={Reviews of Modern Physics},
  volume={43},
  number={1},
  pages={36},
  year={1971},
  publisher={APS}
}

@article{denschlag1998probing,
  title={Probing a singular potential with cold atoms: A neutral atom and a charged wire},
  author={Denschlag, Johannes and Umshaus, Gerhard and Schmiedmayer, J{\"o}rg},
  journal={Physical review letters},
  volume={81},
  number={4},
  pages={737},
  year={1998},
  publisher={APS}
}

@article{plestid2018fall,
  title={Fall to the centre in atom traps and point-particle EFT for absorptive systems},
  author={Plestid, Ryan and Burgess, CP and O’Dell, DHJ},
  journal={Journal of High Energy Physics},
  volume={2018},
  number={8},
  pages={1--39},
  year={2018},
  publisher={Springer}
}

@article{efimov1973energy,
  title={Energy levels of three resonantly interacting particles},
  author={Efimov, V},
  journal={Nuclear Physics A},
  volume={210},
  number={1},
  pages={157--188},
  year={1973},
  publisher={Elsevier}
}

@article{fonseca1979efimov,
  title={Efimov effect in an analytically solvable model},
  author={Fonseca, Antonio C and Redish, Edward F and Shanley, PE},
  journal={Nuclear Physics A},
  volume={320},
  number={2},
  pages={273--288},
  year={1979},
  publisher={Elsevier}
}

@article{braaten2006universality,
  title={Universality in few-body systems with large scattering length},
  author={Braaten, Eric and Hammer, H-W},
  journal={Physics Reports},
  volume={428},
  number={5-6},
  pages={259--390},
  year={2006},
  publisher={Elsevier}
}

@article{moroz2015generalized,
  title={Generalized Efimov effect in one dimension},
  author={Moroz, Sergej and D’Incao, Jos{\'e} P and Petrov, Dmitry S},
  journal={Physical review letters},
  volume={115},
  number={18},
  pages={180406},
  year={2015},
  publisher={APS}
}

@article{calogero1969solution,
  title={Solution of a three-body problem in one dimension},
  author={Calogero, Francesco},
  journal={Journal of Mathematical Physics},
  volume={10},
  number={12},
  pages={2191--2196},
  year={1969},
  publisher={American Institute of Physics}
}

@article{sutherland1971exact,
  title={Exact results for a quantum many-body problem in one dimension},
  author={Sutherland, Bill},
  journal={Physical Review A},
  volume={4},
  number={5},
  pages={2019},
  year={1971},
  publisher={APS}
}

@article{nisoli2014attractive,
  title={Attractive inverse square potential, U (1) gauge, and winding transitions},
  author={Nisoli, Cristiano and Bishop, AR},
  journal={Physical Review Letters},
  volume={112},
  number={7},
  pages={070401},
  year={2014},
  publisher={APS}
}

@article{moroz2010nonrelativistic,
  title={Nonrelativistic inverse square potential, scale anomaly, and complex extension},
  author={Moroz, Sergej and Schmidt, Richard},
  journal={Annals of Physics},
  volume={325},
  number={2},
  pages={491--513},
  year={2010},
  publisher={Elsevier}
}

@article{kaplan2009conformality,
  title={Conformality lost},
  author={Kaplan, David B and Lee, Jong-Wan and Son, Dam T and Stephanov, Mikhail A},
  journal={Physical Review D—Particles, Fields, Gravitation, and Cosmology},
  volume={80},
  number={12},
  pages={125005},
  year={2009},
  publisher={APS}
}

@article{Sundaram:16,
author = {Sriram Sundaram and Prasanta K. Panigrahi},
journal = {Opt. Lett.},
number = {18},
pages = {4222--4224},
publisher = {Optica Publishing Group},
title = {On the origin of the coherence of sunlight on the earth},
volume = {41},
month = {Sep},
year = {2016},
url = {https://opg.optica.org/ol/abstract.cfm?URI=ol-41-18-4222},
doi = {10.1364/OL.41.004222},
}

@article{chamon2011conformal,
  title={Conformal quantum mechanics as the CFT1 dual to AdS2},
  author={Chamon, Claudio and Jackiw, Roman and Pi, So-Young and Santos, Luiz},
  journal={Physics Letters B},
  volume={701},
  number={4},
  pages={503--507},
  year={2011},
  publisher={Elsevier}
}

@article{moroz2010below,
  title={Below the Breitenlohner-Freedman bound in the nonrelativistic AdS/CFT correspondence},
  author={Moroz, Sergej},
  journal={Physical Review D—Particles, Fields, Gravitation, and Cosmology},
  volume={81},
  number={6},
  pages={066002},
  year={2010},
  publisher={APS}
}

@article{birmingham2001near,
  title={Near-horizon conformal structure of black holes},
  author={Birmingham, Danny and Gupta, Kumar S and Sen, Siddhartha},
  journal={Physics Letters B},
  volume={505},
  number={1-4},
  pages={191--196},
  year={2001},
  publisher={Elsevier}
}

@article{camblong2003anomaly,
  title={Anomaly in conformal quantum mechanics: From molecular physics to black holes},
  author={Camblong, Horacio E and Ordonez, Carlos R},
  journal={Physical Review D},
  volume={68},
  number={12},
  pages={125013},
  year={2003},
  publisher={APS}
}

@article{burgess2018effective,
  title={Effective field theory of black hole echoes},
  author={Burgess, CP and Plestid, Ryan and Rummel, Markus},
  journal={Journal of High Energy Physics},
  volume={2018},
  number={9},
  pages={1--31},
  year={2018},
  publisher={Springer}
}

@article{srinivasan1999particle,
  title={Particle production and complex path analysis},
  author={Srinivasan, K and Padmanabhan, T},
  journal={Physical Review D},
  volume={60},
  number={2},
  pages={024007},
  year={1999},
  publisher={APS}
}

@article{daSilva:2025vkl,
    author = "da Silva, U. Camara",
    title = {{Renormalization of Schr{\"o}dinger equation for potentials with inverse-square singularities: generalized trigonometric P{\"o}schl{\textendash}Teller model}},
    eprint = "2503.12715",
    archivePrefix = "arXiv",
    primaryClass = "quant-ph",
    doi = "10.1088/1751-8121/ae2446",
    journal = "J. Phys. A",
    volume = "58",
    number = "50",
    pages = "505201",
    year = "2025"
}

@article{Kunstatter:2008qx,
    author = "Kunstatter, Gabor and Louko, Jorma and Ziprick, Jonathan",
    title = "{Polymer quantization, singularity resolution and the 1/r**2 potential}",
    eprint = "0809.5098",
    archivePrefix = "arXiv",
    primaryClass = "gr-qc",
    doi = "10.1103/PhysRevA.79.032104",
    journal = "Phys. Rev. A",
    volume = "79",
    pages = "032104",
    year = "2009"
}

@article{Bouaziz:2008wxq,
    author = "Bouaziz, Djamil and Bawin, Michel",
    title = "{Singular inverse square potential in arbitrary dimensions with a minimal length: Application to the motion of a dipole in a cosmic string background}",
    eprint = "1009.0930",
    archivePrefix = "arXiv",
    primaryClass = "quant-ph",
    doi = "10.1103/PhysRevA.78.032110",
    journal = "Phys. Rev. A",
    volume = "78",
    pages = "032110",
    year = "2008"
}

@article{Veloso:2025slu,
    author = "Veloso, J. Carvalho and Bakke, K.",
    title = "{Magnetization and Aharonov\textendash{}Bohm effect from the confinement of a point charge to an attractive $r^{-2}$ potential in a uniform magnetic field}",
    doi = "10.1016/j.physb.2025.417033",
    journal = "Physica B",
    volume = "705",
    pages = "417033",
    year = "2025"
}

@article{Veloso:2024ovg,
    author = "Veloso, J. Carvalho and Bakke, K.",
    title = "{Aharonov\textendash{}Bohm effect in an attractive inverse-square potential}",
    doi = "10.1016/j.aop.2024.169902",
    journal = "Annals Phys.",
    volume = "473",
    pages = "169902",
    year = "2025"
}

@article{coleman1973radiative,
  title={Radiative corrections as the origin of spontaneous symmetry breaking},
  author={Coleman, Sidney and Weinberg, Erick},
  journal={Physical Review D},
  volume={7},
  number={6},
  pages={1888},
  year={1973},
  publisher={APS}
}

\end{document}